\documentstyle[aps,pre,epsfig,eqsecnum]{revtex}
\def\SR{{S_R}}
\def\ST{{S_T}}
\def\ut{{\tilde u}}
\def\tu{{v}}
\def\ttu{{\tilde\tu}}
\def\mm#1{\underline{\underline{{#1}}}}

\def\xv{{\mathbf x}}
\def\ev{{\mathbf e}}
\def\Rv{{\mathbf R}}
\def\uv{{\mathbf u}}
\def\nv{{\mathbf n}}
\def\Tr{{\rm Tr}}
\def\um{{\mm{u}}}
\def\vm{{\mm{v}}}

\def\Lm{{\mm{\Lambda}}}

\begin{document}
\twocolumn[\hsize\textwidth\columnwidth\hsize\csname@twocolumnfalse\endcsname
\title{Symmetries and Elasticity of Nematic Gels}
\author{T. C. Lubensky, Ranjan Mukhopadhyay}
\address{Department of Physics, University of Pennsylvania,
Philadelphia, Pennsylvania 19174}
\author{Leo Radzihovsky, Xiangjun Xing}
\address{Department of Physics, University of Colorado,
Boulder, CO 80309}

\date{\today}
\maketitle
\begin{abstract}
  A nematic liquid-crystal gel is a macroscopically homogeneous
  elastic medium with the rotational symmetry of a nematic liquid
  crystal.  In this paper, we develop a general approach to the study
  of these gels that incorporates all underlying symmetries.  After
  reviewing traditional elasticity and clarifying the role of broken
  rotational symmetries in both the reference space of points in the
  undistorted medium and the target space into which these points are
  mapped, we explore the unusual properties of nematic gels from a
  number of perspectives.  We show how symmetries of nematic gels
  formed via spontaneous symmetry breaking from an isotropic gel
  enforce soft elastic response characterized by the vanishing of a
  shear modulus and the vanishing of stress up to a critical value of
  strain along certain directions. We also study the phase transition
  from isotropic to nematic gels. In addition to being fully
  consistent with approaches to nematic gels based on rubber
  elasticity, our description has the important advantages of being
  independent of a microscopic model, of emphasizing and clarifying
  the role of broken symmetries in determining elastic response, and
  of permitting easy incorporation of spatial variations, thermal
  fluctuations, and gel heterogeneity, thereby allowing a full
  statistical-mechanical treatment of these novel materials.
\end{abstract}
\pacs{PACS: } \vskip2pc]

\section{Introduction}

The term liquid crystal\cite{deGennesProst93,Chandrasekhar92} has
traditionally been used to describe phases of matter that exhibit
anisotropies characteristic of crystals but that under appropriate
conditions flow like a liquid.  These phases typically have
symmetries intermediate between that of a homogeneous isotropic
fluid and that of a three-dimensional periodic crystalline solid.
Indeed one can provide an almost complete characterization of a
liquid-crystalline phase by specifying its symmetry.  For
example, the nematic phase, which is spatially homogeneous yet
optically uniaxial has $D_{\infty h}$ symmetry.  The typical
phase sequence for a thermotropic liquid crystal on cooling
begins with an isotropic fluid and ends with a crystalline solid
after passing through nematic, layered smectic-$A$ and
smectic-$C$, and possibly hexatic phases.

There is, however, a large variety of materials that have the same
macroscopic symmetry as fluids, but that cannot flow: they are
macroscopically homogeneous and isotropic elastic media with a
nonvanishing shear modulus that provides resistance to shear
distortions. We will refer to these materials, which include
everything from glasses to elastomers or
rubbers\cite{comment_glassOP}, as gels\cite{glass_gel}.  One can
imagine phases arising from a reference state of a {\em gel}
(rather than a liquid) with the same macroscopic symmetries as
conventional liquid crystals.  As we shall discuss more fully
below, these phases do in fact
exist\cite{FinKoc81}-\cite{Leheny}, and, because they cannot
flow, they have mechanical properties and mode structures that
differ significantly from those of standard liquid crystals. We
will call these phases ``liquid-crystal gels'' because they are
{\em gels} with the symmetry of conventional (fluid) liquid
crystals. In this paper, we will develop a powerful and general
formalism to describe nematic gels and use it to explore their
remarkable properties. We will focus particularly on nematic gels
that form via spontaneous orientational symmetry breaking from an
isotropic gel phase.  Our formalism can be generalized to treat
other liquid-crystalline gel phases.

There are a large number of experimental realizations of
liquid-crystal gels. Of particular interest to us are
liquid-crystal elastomers\cite{FinKoc81}-\cite{Terentjev99}. These
materials, which are formed by weakly crosslinking either
side-chain\cite{FinKoc81} or main-chain\cite{Zentel89} polymers,
combine the enormous extensibility of rubbers with the
orientational properties of liquid crystals.  They are, therefore,
of considerable technological importance. The existence of the
rubbery crosslinked network appears to have relatively little
effect on liquid-crystalline phase behavior, and the standard
thermotropic nematic, cholesteric, smectic-$A$, and smectic-$C$
phases have their elastomeric
counterparts\cite{Terentjev99,KimFin01}. The elastic properties of
these phases do, however, crucially depend on whether a given
liquid-crystalline order was established before or after
crosslinking.  Liquid-crystal gels can also be prepared in other
ways, for example by polymerization of monomer solutes in a
liquid-crystalline solvent\cite{Bowman97}, or by confining
conventional liquid crystal inside a dilute flexible matrix, such
as, e.g., aerosil\cite{CrawfordZumer96,Leheny}.

To fully characterize liquid-crystal gel phases, two complementary
basic questions must be addressed: (1) What effect does
liquid-crystal order have on the gel {\em elasticity}? (2) How
does the rigidity of the underlying gel affect liquid-crystal
{\em order} and its {\em stability} to fluctuations?  Our primary
concern in this paper will be with question (1) applied to
nematic gels. With regard to question (2), gel elasticity of
weakly crosslinked elastomers has relatively little effect on the
existence of liquid-crystalline phases. On the other hand,
liquid-crystal elastomers appear experimentally to be more
strongly ordered than their fluid counterparts. For example unlike
conventional nematics, which are milky and therefore have domain
sizes at the scale of visible light, elastomer nematics are clear,
indicating orientational order extending beyond a micron. This
property directly and clearly follows from our model of nematic
elastomers, as well as early work\cite{Olmsted94}, and it will be
explored in more detail in a future publication.\cite{unpublished}

\begin{figure}
  \centerline{\epsfbox{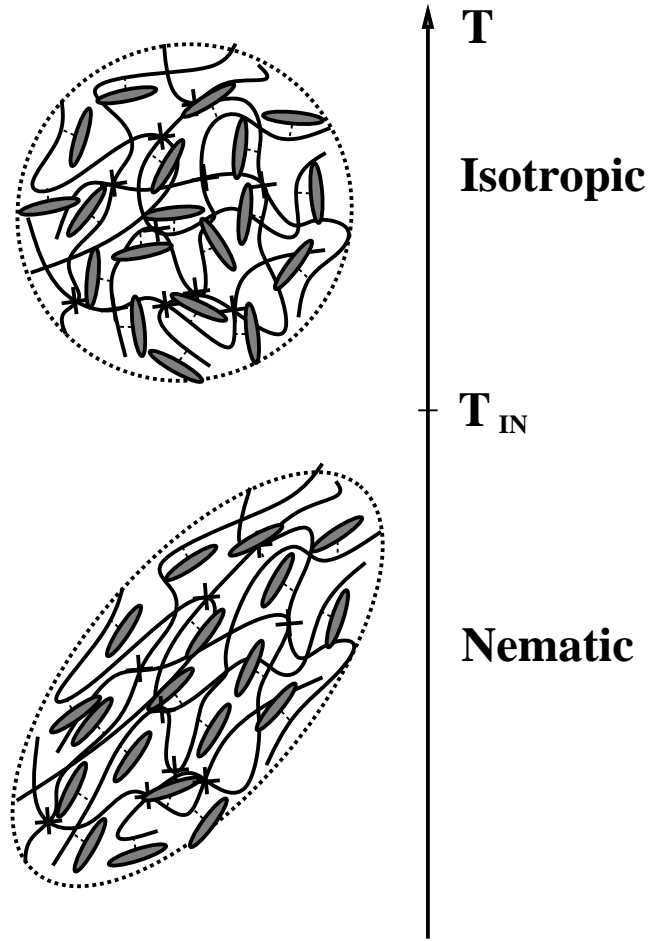}}
  \caption{A cartoon of a liquid-crystal gel undergoing an
    Isotropic-Nematic transition, accompanied by a spontaneous
    uniaxial distortion.}
\label{elastomerIN}
\end{figure}

The strong interplay between broken symmetry and the nature of
long-wavelength excitations of ordered phases is a major theme of
physics\cite{ChaLub95}.  Symmetry principles dictate that ordered
thermodynamic phases that break a continuous symmetry have
low-energy distortions, or ``soft" modes, that are described by an
elastic energy depending only on gradients of these Goldstone
fields whose spatially uniform increments take the system to
symmetry-equivalent states.  The form of this elastic free energy
is uniquely determined by the properties of the reference phase
whose symmetry is broken and by the nature of the broken symmetry
itself.  Conventional nematic liquid crystals break the rotational
isotropy of an isotropic homogeneous liquid, and they are
characterized by the Frank elastic free
energy\cite{deGennesProst93,Chandrasekhar92}, which is a
functional of the Goldstone field ${\bf n}$, the Frank director
specifying the direction of molecular alignment. As illustrated in
Fig.\ \ref{elastomerIN}, nematic phases of liquid-crystal gels
that form from an isotropic gel state also spontaneously break
rotational isotropy.  Their long-wavelength elastic energy,
however, differs significantly from the Frank free energy of
conventional nematics because the reference gel state (unlike a
reference fluid state) has a nonvanishing shear modulus.  The
elastic energy of a spontaneously formed nematic gel was first
calculated by Golubovi\'{c} and Lubensky (GL) \cite{GolLub89} in
their study of a model isotropic elastic medium that undergoes a
phase transition to a uniaxial state when its shear modulus
becomes smaller than a critical value.  They found that the
Goldstone fields of a uniaxial gel are displacement fields and
that their associated elastic energy is expressed as a function of
standard strains of a solid. Normally, the elastic energy of a
uniaxial elastic medium is characterized by 5 independent elastic
constants.  A nematic gel that forms spontaneously from an
isotropic gel is significantly softer than a conventional uniaxial
solid: it is characterized by ``soft" elasticity in which the
elastic constant $C_5$ associated with shears in the plane
containing the anisotropy axis vanishes and in which stress
vanishes up to critical values of certain
shears\cite{FinKun97}-\cite{Warner99}.

Many of the properties of nematic elastomers can be explained by
an elegant and remarkably simple extension\cite{BlaTer94} of
standard rubber elasticity\cite{Treolar75} in which nematic order
leads to an anisotropic step-length tensor for random-flight
polymer segments between crosslinking points. This
``neoclassical'' theory of rubber elasticity, which describes a
particular realization of an anisotropic gel exhibits ``soft"
elasticity\cite{WarBla94,Olmsted94} in accord with the general
symmetry-based predictions of GL.

In this largely pedagogical paper, we will explore the elastic and
orientational properties of nematic gels from a perspective that is an
extension of that of Ref.\ \onlinecite{GolLub89} rather that of rubber
elasticity\cite{WarTer96}. In particular we will describe elastic
properties mostly in terms of nonlinear strain tensors familiar from
the elastic theory of solids and membranes\cite{LandauEL,membrane}
rather than the perhaps more fundamental Cauchy deformation tensor
used in rubber elasticity\cite{Treolar75} from which the nonlinear
strain tensors can be constructed.  We will usually measure strain
relative to some equilibrium state in which the strain is zero rather
than relative to a state at the time of preparation as is common
practice in the theory of (incompressible) rubbers. Our approach based
on nonlinear strains that are invariant under arbitrary rotations of
the sample (or, alternatively, as we shall see, the reference state)
allows us to keep track of rotational invariances with relative ease.
It is particularly well suited, as we will show in a future
publication\cite{membraneXing}, to the treatment of renormalized
elasticity arising from the interplay of thermal fluctuations and
nonlinear elasticity.  The approach is also convenient for the
discussion of external-field-induced instabilities of an equilibrium
phase.  Most importantly, our formalism elucidates the origin of the
novel soft elasticity of nematic gels, making it clear that it arises
from general symmetry principles common to any {\em spontaneously}
uniaxially ordered elastic medium and is {\em not} limited to any
specific model of such materials. Our description has the disadvantage
compared to the rubber elasticity approach that it does not naturally
treat the very large (as much as $400\%$\cite{FinWer00}) extensions
that can arise in elastomers.

Though gels are macroscopically isotropic and homogeneous, they
are always randomly anisotropic and inhomogeneous at sufficiently
short length scales.  Consequently, there is a local preferred
direction of orientational (and spatial) order that acts as a
random orienting (and pinning) field.  These quenched fields are
certainly present in anisotropic gels: optical observations in
thin films provide direct evidence of their
existence\cite{ChaChi97,ClaNis97}.  One consequence of such
random quenched local fields is that elastomers crosslinked in
the isotropic phase and cooled into the nematic phase exhibit a
polydomain orientational structure, which disappears at a
polydomain-monodomain transition when a sufficiently large
external stress is applied\cite{FriTer99}.  Such random static
fields can easily be incorporated in our formulation of nematic
gels. Their study is in principle necessary to understand
completely the effect of gel matrix on nematic order [question
(2) above]. Such an investigation would parallel a body of work
on conventional liquid crystals confined inside the quenched,
random, but (nearly) nondeformable environment of rigid gels,
such as e.g., an aerogel.\cite{CrawfordZumer96,RTaerogel,Feldman}
Experience with these rigid systems indicate that random fields
might become qualitatively important at sufficiently long
scales.  Nevertheless, in this article, we will ignore completely
effects of random fields and concentrate on properties of
anisotropic gels formed from ideal isotropic homogeneous gels. By
focusing on gels in which crosslinks are dense and
well-percolated, we will also have nothing to say about the
nature of the vulcanization transition itself\cite{Goldbart}.

We will also leave for the future\cite{semi-soft} the analysis of
``semi-soft" nematic elastomers\cite{FinKun97,Warner99,VerWar95} that
are prepared by polymer crosslinking in the nematic phase.  These
materials are characterized by a small non-vanishing elastic modulus
$C_5$ and nonlinear stress-strain curves with a small but nonvanishing
stress up to large strains.

This paper is organized as follows.  Section II reviews the
standard Lagrangian theory of elasticity and establishes notation
for sections that follow.  It introduces the reference space
consisting of points in the undistorted medium and the target
space into which these points are mapped. This section emphasizes
the {\em two} distinct rotational invariances of isotropic elastic
media, namely invariance with respect to rotation of the deformed
sample itself (rotations in the target space) and invariance with
respect to rotation of points in the original reference material
that map to particular points in the target space (rotations in
the reference space). Section II also discusses the standard
nonlinear strain tensor, the right Cauchy-Green strain tensor,
that is invariant with respect to rotations in the target space
and introduces an alternative nonlinear strain tensor, the left
Cauchy-Green strain tensor, that is invariant with respect to
rotations of the reference space but transforms like rank-2 tensor
in the target space. Section III elaborates on the model
considered in Ref.\ \onlinecite{GolLub89}.  It shows in particular
that the elastic energy of the anisotropic gel phase expanded only
to harmonic order in the nonlinear strain does not preserve the
rotational invariance of the original energy with respect to
rotations in the reference space. It also discusses the
isotropic-to-anisotropic transition in terms of the alternative
strain tensor that preserves rotational invariance in the
reference space. Section IV discusses a model with both strain and
the symmetric-traceless tensor order parameter $Q_{ij}$ of the
nematic state.  It shows that the ``soft'' elasticity of the
anisotropic (nematic) state arises because the nematic order
parameter can relax strains in the plane containing the anisotropy
axis, as first shown by Olmsted\cite{Olmsted94}. Section IV also
derives the elastic energy for nematic glasses deep in the ordered
phase, where biaxial fluctuations can be neglected and nematic
properties can be described completely by the Frank director. This
theory is expressed in terms of generalized nonlinear strains that
are functions of strain and relative director orientation and that
are invariant with respect to arbitrary simultaneous rotations of
the director and mass points. Section V explores the relation
between the theory presented here and the neoclassical elastomer
theory. Section VI concludes with a discussion of some of the many
interesting open problems, such as instabilities of elastomers
induced by external perturbations (e.g., electric or magnetic
fields) and the effects of thermal fluctuations and quenched local
random anisotropy fields, that can conveniently be addressed
through our formulation

\section{Classical Lagrangian Elasticity}
\label{Elasticity}

Classical elasticity\cite{LandauEL} provides a phenomenological
description of the energy associated with
slowly\cite{comment_slow} varying distortions of an elastic body
from its equilibrium configuration.  As discussed in the
Introduction, it is a symmetry-restricted theory of the low-energy
Goldstone modes associated with spontaneous translational symmetry
breaking. In this section, we will review the classical theory of
Lagrangian elasticity\cite{LagrangeE}, introducing concepts that
will be important for our study of spontaneously uniaxial nematic
elastomers.

\subsection{Strain}
\label{strain}
The equilibrium unstretched medium occupies a region of a
Euclidean 3-space, which we will call the reference space $\SR$.
Mass points in this medium are indexed by their vector positions
$\xv=(x_1,x_2,x_3)\equiv(x,y,z)$ in $\SR$, which are their
positions in the unstretched medium.  When the medium is
distorted, the point originally at $\xv$ is mapped to a new point
$\Rv( \xv )= (R_1(\xv),R_2(\xv),R_3(\xv))$ in Euclidean space.  We
will refer to the space of points defined by $\Rv$ as the target
space $\ST$. Since there is no distortion when $\Rv(\xv) = \xv$,
it is useful to introduce the displacement vector $\uv( \xv )$
that measures the deviation of $\Rv$ from $\xv$:
\begin{equation}
\Rv(\xv) = \xv + \uv( \xv) .
\end{equation}
Both $\SR$ and $\ST$ are Euclidean, with distances determined by the
unit metric: $dx^2 = dx_i dx_i$, and $dR^2 = dR_i dR_i$, where the
Einstein summation convention on repeated indices is
understood.\cite{two-spaces_comment} It is often interesting to
consider generalizations of the above picture to a $D$-dimensional
reference space and a $d\geq D$ target space, for example to describe
$D=2$-dimensional tethered membranes fluctuating in $d=3$-dimensional
real space\cite{membrane}.  In this paper, however, we will restrict
our attention to $D=d=3$, leaving discussion of membranes to a future
publication.\cite{membraneXing}

Distortions that vary slowly on a scale set by microscopic lengths of
the reference material (interparticle separation in a glass, distance
between crosslinks in an elastomer, etc.) are described by the Cauchy
deformation tensor\cite{two-spaces_comment,strain},
\begin{equation}
\Lambda_{ij} = {\partial R_i \over \partial x_j } \equiv
\partial_j R_i =  \delta_{ij} + \eta_{ij} ,
\label{eq:Lam1}
\end{equation}
where
\begin{equation}
\eta_{ij} = \partial_j u_i\
\label{eq:st-def}
\end{equation}
is the displacement gradient tensor. Throughout the paper, we will
often use matrix notation in which $\mm{M}$ is the matrix with
components $M_{ij}$ and $\mm{M}^T$ is the transpose matrix with
components $M_{ji}$.

The energy of the distorted state relative to the undistorted
one depends on the how much the target space is stretched
relative to the reference space, i.e., by how much the
distance between two nearby points changes in the mapping
from the reference to the target:
\begin{equation}
dR^2 - dx^2 = 2 u_{ij} dx_i dx_j ,
\label{length}
\end{equation}
where
\begin{eqnarray}
u_{ij} &= &\case{1}{2}\left( \Lambda_{ki} \Lambda_{kj} -
\delta_{ij} \right )\qquad  {\rm or} \qquad \mm{u} =
\case{1}{2}\left(\mm{\Lambda}^T \mm{\Lambda} -
\mm{\delta}\right) \nonumber\\
& =& \case{1}{2}(\partial_i u_j + \partial_j u_i +
\partial_i u_k \partial_j u_k )\nonumber\\
& = & \case{1}{2}(\eta_{ij}+ \eta_{ji} + \eta_{ki} \eta_{kj}) .
\label{eq:strain1}
\end{eqnarray}
$u_{ij}$ is the familiar nonlinear Lagrangian strain
tensor\cite{LandauEL}, also called the {\em right} Cauchy-Green
strain tensor or simply the Green strain tensor\cite{strain}. It
is symmetric by construction. It is also {\em invariant} (i.e.,
transforms as a {\em scalar}) under arbitrary rotations of the
{\em target} space vector $\Rv$, i.e., if $R_i$ is replaced by
$R_i' = O_{T ij} R_j$, where $O_{T ij}$ is an arbitrary rotation
matrix, $u_{ij}$ does not change.  On the other hand, $u_{ij}$
transforms like a rank-2 {\em tensor} under rotations of the {\em
reference} space, i.e., if $x_i \rightarrow x_i' = O^{-1}_{R ij}
x_j$, then\cite{inverse_note}
\begin{equation}
\mm{u} \rightarrow \mm{O}_R \mm{u}\ \mm{O}_R^{-1} .
\end{equation}
Isotropic solids, e.g., the glasses and gels of interest to us, are
(statistically) invariant under arbitrary rotation $\mm{O}_R$ in the
reference space $S_R$. Crystals, on the other hand, have lower
symmetry and are invariant only under a point subgroup of all
rotations $\mm{O}_R$.

In contrast, invariance with respect to arbitrary rotations in $\ST$
is a property of {\em all} elastic media in the absence of external
aligning fields, whether they be isotropic, crystalline, or wildly
inhomogeneous. Thus, because it by construction incorporates the $O_T$
invariance, in most instances, $u_{ij}$ is the strain tensor in terms
of which elastic theory is most conveniently formulated.  However,
here we are interested in systems (gels) that exhibit rotational
invariance in the {\em reference} space, i.e., an $O_R$ invariance of
$S_R$, and, therefore, a distinct {\em left} Cauchy-Green strain
tensor,
\begin{eqnarray}
\tu_{ij} &= &\case{1}{2}\left( \Lambda_{ik} \Lambda_{jk} -
\delta_{ij} \right) \,\, {\rm or} \,\, \mm{\tu} =
\case{1}{2}\left(\Lm \Lm^T - \mm{\delta}\right)\nonumber\\
& = & \case{1}{2}(\partial_i u_j + \partial_j u_i +
\partial_k u_i \partial_k u_j ) \nonumber \\
& = & \case{1}{2}(\eta_{ij} + \eta_{ji} + \eta_{ik}\eta_{jk}),
\label{eq:strain2}
\end{eqnarray}
is useful\cite{deformations}.  This tensor is invariant under
arbitrary rotations $\mm{O}_R$ in $S_R$, but it transforms like a
rank-2 tensor under rotations $\mm{O}_T$ in $S_T$:
\begin{equation}
\mm{v} \rightarrow \mm{O}_T\mm{v}\ \mm{O}_T^{-1} .
\end{equation}
In what follows, we will simply refer to $\mm{u}$ and $\mm{v}$ as
right and left strain tensors, respectively.

The left strain tensor $\mm{v}$ can be contracted with other
target-space tensors, such as the Maier-Saupe-de Gennes nematic order
parameter $Q_{ij}$, or the electric field $E_i$, to form scalar
invariants such as ${\rm Tr} \mm{v}\mm{Q}$ or $E_i v_{ij} E_j$. In
contrast, the contractions ${\rm Tr}\mm{u}\mm{Q}$ and $E_i u_{ij} E_j$
are {\em not} scalars since $\mm{u}$ does not transform like a tensor
in the same space as the tensors $Q_{ij}$ and $E_i E_j$.  In the
absence of external aligning fields such as ${\bf E}$ that effectively
render the target space anisotropic, the right strain $\mm{u}$
provides a complete description of elastic distortions, even if, as is
the case for crystals, the reference space is anisotropic. If $S_R$ is
isotropic and there are external fields breaking the isotropy of
$S_T$, then $\mm{u}$ cannot provide a similar complete description,
but the left strain $\mm{v}$ can. On the other hand, the left strain
cannot provide a complete description if the reference space is not
isotropic. For example, semi-soft elastomers crosslinked in the
nematic phase with a director ${\bf n}_0$, which specifies a direction
in $S_R$, are invariant under the simultaneous rotations of $\nv_0$
and $\xv$, ${\bf n}_0 \rightarrow \mm{O}_R^{-1} {\bf n}_0$ and ${\bf
  x} \rightarrow \mm{O}_R^{-1} {\bf x}$, but not under rotations,
${\bf x} \rightarrow \mm{O}_R^{-1} {\bf x}$, of $\xv$ alone. The left
strain $\mm{v}$ is a scalar in $S_R$, and it cannot be contracted with
the reference space vector $\nv_0$. Thus, it is impossible to
construct scalar invariants involving $\mm{v}$ and $\nv_0$ and to
construct a free energy in terms of $\mm{v}$ that reflects the
anisotropy of $S_R$. If $S_R$ is anisotropic and there are external
fields breaking the rotational invariance of $S_T$, then only the
deformation tensor $\mm{\Lambda}$ can provide a complete description
of the energy of elastic distortions.

\subsection{Isotropic systems}
\label{iso-strain}

\begin{figure}
  \centerline{\epsfbox{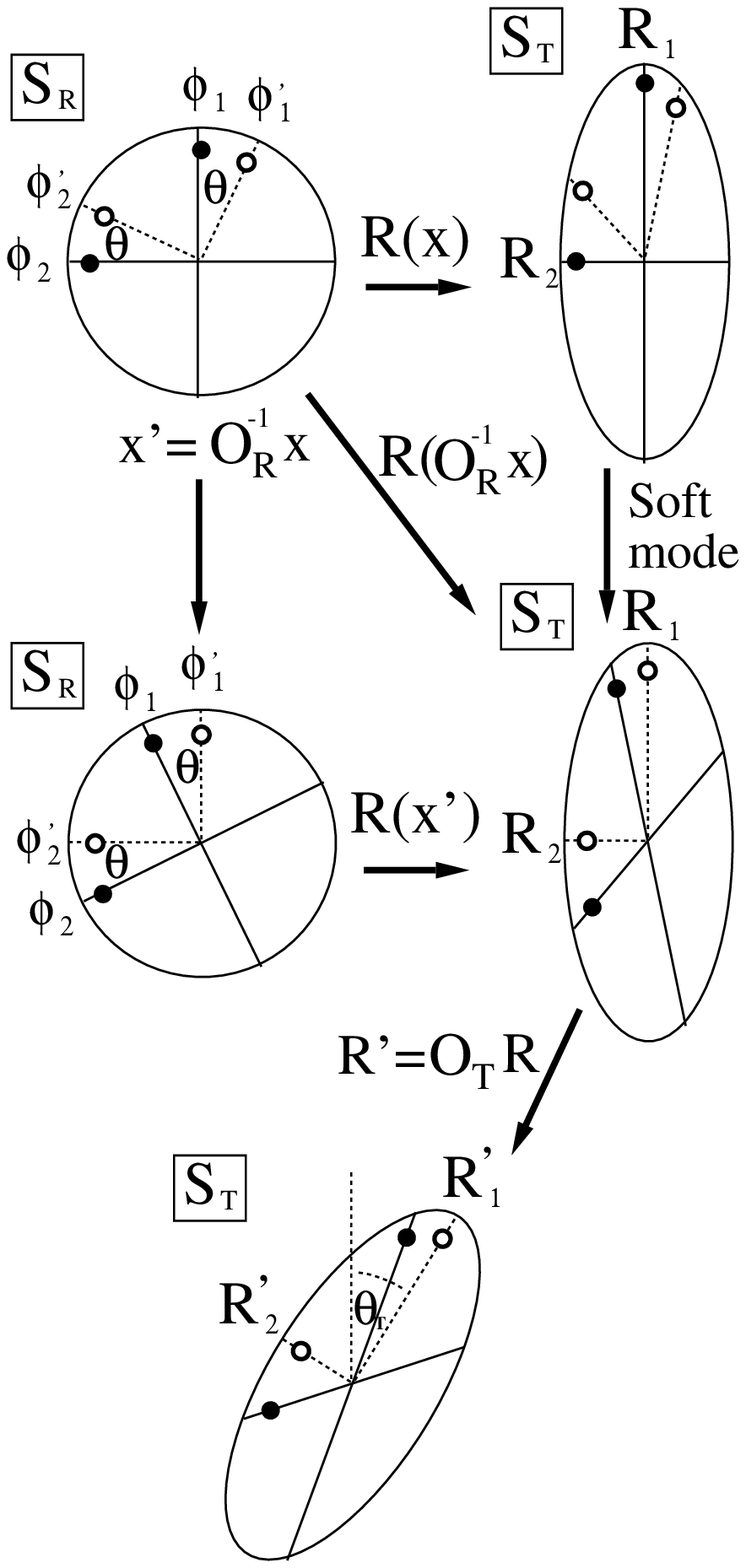}}
  \caption{Schematic representation of mappings from the reference
    space $S_R$ to the target space $S_T$.  The points
    $\xv_1=(r,\phi_1)$ and $\xv_2=(r,\phi_2)$ in $S_R$ are mapped,
    respectively, to the points $\Rv_1$ and $\Rv_2$ in $S_T$. There is
    a strain energy $E_S$ associated with this mapping.  For isotropic
    reference spaces, mapping of points in $S_R$, first rotated by
    $\theta$ inside the reference space (described by a rotation
    matrix $\mm{O}^{-1}_R$) to the same set of points in $S_T$, i.e.,
    mapping points $\xv_1' = (r, \phi_1 - \theta)$ and $\xv_2'=(r,
    \phi_2 - \theta)$ to $\Rv_1$ and $\Rv_2$ clearly produces the same
    energy $E_S$ as the unrotated mapping.  Subsequent target-space
    rotation $\mm{O}_T$ by $\theta_T$ of the resulting distorted state
    with points $\Rv_1$ and $\Rv_2$ mapped to $\Rv_1'$ and $\Rv_2'$
    costs no energy. The transformation $\Rv(\xv) \rightarrow
    \mm{O}_T\Rv(\mm{O}_R^{-1} \xv )$ between energetically-equivalent
    nematic states is the Goldstone mode responsible for the novel
    elastic properties of nematic elastomers.}
\label{fig1}
\end{figure}

For most gels, the reference space is macroscopically isotropic
and homogeneous, i.e., like an isotropic fluid, it is invariant
under $\xv \rightarrow {\bf T}+ \mm{O}_R^{-1} \xv$ for arbitrary
translations ${\bf T}$ and rotations $\mm{O}_R$ in $S_R$. Thus,
the elastic energy is invariant under $\Rv ( \xv) \rightarrow
\mm{O}_T\Rv({\bf T} + \mm{O}_R^{-1} \xv )$.  The invariance under
rotations $\mm{O}_T$ of the target space is easy to understand:
different physical orientations of the material (even if
arbitrarily distorted) have the same energy. Invariance under
$\mm{O}_R$ is somewhat more subtle though complementary. Figure
\ref{fig1} provides a useful graphic representation of this
invariance in two dimensions.  Consider a circle of radius $r$ in
the reference space consisting of the points $\xv = r (\cos \phi,
\sin \phi ) \equiv (r,\phi)$. Under distortion, it is mapped onto
some closed curve in $\ST$ consisting of points $\Rv ( \phi )$.
Thus, the point $(r,\phi_1)$ in $\SR$ is mapped to the point
$\Rv_1 = \Rv (\phi_1)$ in $\ST$, $(r,\phi_2)$ is mapped to $\Rv_2
= \Rv (\phi_2)$, and so on.  Under a rotation through $\theta$ in
$\SR$, $\phi \rightarrow \phi + \theta$.  Because of the isotropy
of the undistorted, reference state, the energy is not changed if
the points $(r,\phi-\theta)$ rather than $(r,\phi)$ are mapped to
the points $\Rv(\phi)$, i.e., if $(r,\phi_1 - \theta )$ is mapped
to $\Rv_1$, $(r, \phi_2 - \theta )$ to $\Rv_2$ and so on.

Care must be taken to incorporate the above symmetries in the free
energy density of such homogeneous and isotropic gels.  Invariance
with respect to translations ${\bf T}$ in $S_R$ is enforced by
requiring that the free energy density depend only on spatial
derivatives of $\Rv$ with respect to $\xv$, i.e., depend only on
$\mm{\Lambda}$ and possibly higher derivatives of $\Rv$.  Under
rotations in $S_R$ and $S_T$, the Cauchy strain tensor
$\Lambda_{ij}$ transforms according to
\begin{equation}
\Lambda_{ij} \rightarrow O_{T ik} {\partial R_k \over \partial
x'_l} {\partial x'_l\over \partial x_j} = O_{T ik}
\Lambda_{kl}O^T_{R lj} .
\end{equation}
The free energy density $f$ of an isotropic gel is invariant under
{\em independent} $\mm{O}_T$ and $\mm{O}_R$ rotations and must
satisfy
\begin{equation}
f(\mm{\Lambda}) = f(\mm{O}_T\mm{\Lambda}\ \mm{O}_R^{-1} ) .
\end{equation}
It must, therefore, be constructed from the scalar invariants ${\rm
  Tr}(\mm{\Lambda} \mm{\Lambda}^T)^n$ and
$\det\mm{\Lambda}\mm{\Lambda}^T= (\det \Lambda )^2$\cite{invar}.
Alternatively, the free energy can be equivalently expressed in terms
of $\mm{u}$ or $\mm{v}$, with the respective invariances
\begin{mathletters}
\begin{eqnarray}
f(\mm{u}) & = & f(\mm{O}_R\mm{u}\ \mm{O}_R^{-1}) \label{fu}\\
f(\mm{v}) & = & f(\mm{O}_T\mm{v}\ \mm{O}_T^{-1}) \label{fv},
\end{eqnarray}
\end{mathletters}
which are enforced by allowing only fully contracted powers of
strain tensors to appear. The energies $f(\mm{u})$ and
$f(\mm{v})$ can be derived from $f(\mm{\Lambda})$ using
\begin{eqnarray}
\det\mm{\Lambda} \mm{\Lambda}^T &= &\exp{\rm Tr} \ln (\mm{\delta}
+ 2 \mm{u}) \nonumber \\
& =& \exp {\rm Tr} \ln (\mm{\delta} + 2 \mm{v}) , \nonumber\\
\Tr\um^n & = & \Tr \left[\case{1}{2} \left(\Lm^T \Lm
-\mm{\delta}\right)\right]^n = \Tr
\left[\case{1}{2}\left(\Lm\Lm^T-\mm{\delta}\right)\right]^n
\nonumber \\
& = & \Tr \mm{v}^n .
\label{uv}
\end{eqnarray}
Thus, $f(\mm{u})$ and $f(\mm{v})$ depend only on ${\rm Tr}\mm{u}^n
= {\rm Tr} \mm{v}^n$, and $f(\mm{v})$ is the {\em same} function
of $\mm{v}$ that $f(\mm{u})$ is of $\mm{u}$. For the discussion of
encoding these two $O_R$ and $O_T$ symmetries at the harmonic
level in the phonon variable ${\bf u}$, see Appendix \ref{app:C}.

Although many of the properties of nematic elastomers follow
directly from the above invariances, in what follows, it will be
useful to have explicit forms for the elastic free energy density.
A model free energy in terms nonlinear strain tensor $\um$, up to
fourth order in $\um$ is
\begin{eqnarray}
f(\mm{u})& = & \case{1}{2} \lambda (\Tr \um)^2 + \mu \Tr \um^2 -
C \Tr \um^3 \nonumber \\
& & + D' (\Tr \um^2 )^2 - E' \Tr \um \Tr \um^2 .
\label{f_1}
\end{eqnarray}
As just discussed, this free energy can equally well be expressed in
terms of $\mm{v}$ merely by replacing $\um$ by $\mm{v}$.  Invariances
with respect to rotations in $S_R$ and $S_T$ are enforced in $f(\um)$
in different ways. Symmetry under $\mm{O}_T$ is enforced by the
construction of the strain tensor $\um$, Eq.\ (\ref{eq:strain1}),
which, being a scalar in $S_T$, is automatically invariant under
$\mm{O}_T$. Invariance under $\mm{O}_R$ is enforced by only allowing
terms in $f(\um)$ that transform as a scalar under $\mm{O}_R$, i.e.,
only fully contracted powers of $\um$.  In contrast, invariance of
$f(\mm{v})$ with respect to $\mm{O}_R$ is enforced by construction of
the strain tensor $\vm$ (a scalar in $S_R$), whereas that with respect
to $\mm{O}_T$ is enforced by only allowing terms in $f(\mm{v})$ that
transform like a scalar under $\mm{O}_T$, i.e., requiring that all the
target space indices be contracted.

As usual, the reference state, relative to which $\um$ is defined is
taken to be in mechanical equilibrium, guaranteeing that no terms
linear in $\um$ appear. The first two terms of $f$ are the standard
elastic energy of an isotropic medium with $\lambda$ and $\mu$ the
Lam\'{e} coefficients\cite{LandauEL}.  We have included stabilizing
nonlinear terms in the strain tensor $\um$ because we will eventually
want to consider phase transitions to an anisotropic state induced by
a decrease in the shear modulus $\mu$ below a critical value. In the
spirit of Landau theory of phase transitions\cite{ChaLub95}, at
present, we view $\mu$ as a phenomenological parameter that is allowed
to vary and even become negative.  As we shall see in more detail in
Sec.\ \ref{nemgels}, the origin of a diminishing $\mu$ in
liquid-crystal elastomers is the instability of the isotropic state
toward the development of nematic liquid-crystal order characterized
by the Maier-Saupe order parameter $\mm{Q}$.  In $f$, we have left out
one third-order and three fourth-order terms permitted by symmetry,
namely ones proportional to $(\Tr \um)^3$, $(\Tr \um)^4$, $(\Tr \um)^2
\Tr \um^2$ and $\Tr \um \Tr \um^3$, respectively.  Though these terms
can easily be included, their effect is small for the nearly
incompressible systems of most interest to us.

Our primary interest is in the state with {\em spontaneously} broken
rotational symmetry that is produced when $\mu$ falls below a critical
value.  To describe this state and the transition to it, it is useful
to decompose $u_{ij}$ into its scalar (in $S_R$) and
symmetric-traceless parts:
\begin{equation}
u_{ij} = \case{1}{3} \delta_{ij} u_{kk} + \ut_{ij} ,
\label{ut}
\end{equation}
where
\begin{equation}
\ut_{ij} = u_{ij} - \case{1}{3} \delta_{ij} u_{kk} .
\label{eq:st-strain}
\end{equation}
Using Eq.\ (\ref{ut}) in Eq.\ (\ref{f_1}) and keeping only the
lowest order terms in $\Tr \um$, we obtain the model free energy
density that we will use in discussions of the anisotropic state
and the transition to it:
\begin{equation}
f  = \case{1}{2} B [\Tr \um -(E/B)\Tr \mm{\ut}^2]^2 + f_1
\label{eq:f}
\end{equation}
with
\begin{equation}
f_1 = \case{1}{2}A \Tr \mm{\ut}^2 - C \Tr\mm{\ut}^3 + D (\Tr
\mm{\ut}^2)^2 ,
\label{eq:f_2}
\end{equation}
where $A = 2\mu$, $B = \lambda + \case{2}{3} \mu$ is the bulk
modulus, $E= E' - C$, $D=D'-E^2/(2B)$ and for simplicity we have
dropped qualitatively inconsequential cubic and quartic terms in
$\Tr\um$.\cite{incompressible_comment}

\subsection{Anisotropic systems}

Often the reference state is a crystal that is invariant only
under operations of some subgroup of $O_3$.  In this case, there
are additional combinations of the strain tensor that are
invariant under the reduced set of symmetry operations of $S_R$,
and the elastic energy is in general described, to harmonic order
in $u_{ij}$, in terms of a 4th rank elastic-constant tensor
$C_{ijkl}$, with $f=\case{1}{2}C_{ijkl}u_{ij}u_{kl}$.  We will be
particularly interested in uniaxial systems with axis along
${\mathbf{n}}_0$, for which the general form of the
elastic-constant tensor is (but see Sec.\ \ref{strainonly})
\begin{eqnarray}
C_{ijkl}&=&C_1 n_{0i} n_{0j} n_{0k} n_{0l}+ C_2
(n_{0i}n_{0j}\delta_{kl}^{0\perp}+n_{0k}n_{0l}\delta_{ij}^{0\perp})\nonumber\\
& +&  C_3\delta_{ij}^{0\perp}\delta_{kl}^{0\perp} +
C_4(\delta_{ik}^{0\perp}\delta_{jl}^{0\perp} +
\delta_{il}^{0\perp}\delta_{jk}^{0\perp})  \\
&+ &  \case{1}{2} C_5(\delta_{ik}^{0\perp} n_j^0 n_l^0 +
\delta_{il}^{0\perp} n_j^0 n_k^0 +\delta_{jk}^{0\perp} n_i^0 n_l^0
+ \delta_{jl}^{0\perp} n_i^0 n_k^0 ), \nonumber
\end{eqnarray}
where $\delta_{ij}^{0\perp} = \delta_{ij} - n_{0i} n_{0j}$. The
elastic energy in three dimensions with the $z$-axis chosen along
${\mathbf{n}}_0$ is
\begin{eqnarray}
f_{\rm uni} &= &\case{1}{2} C_1 u_{zz}^2 +
C_2 u_{zz} (u_{xx} + u_{yy} ) + \case{1}{2}C_3 (u_{xx} + u_{yy})^2 \nonumber \\
& & + C_4 (u_{xx}^2 + u_{yy}^2 + 2 u_{xy}^2) + C_5 (u_{xz}^2 +
u_{yz}^2 ) .
\label{funi}
\end{eqnarray}
The strain $u_{ij}$ is still invariant under arbitrary rotation in
$S_T$, so $f_{\rm uni}$ is invariant under these rotations, as it must
be.  The reduced symmetry of the reference state introduces an
asymmetry between the reference and target spaces, and it is no longer
so useful to introduced the alternative strain tensor $\mm{\tu}$
unless we wish to discuss explicitly coupling between strain and
another target-space tensor-field order parameter, such as the
Maier-Saupe order parameter for a nematic.

The elastic energy $f_{\rm uni}$ of Eq.\ (\ref{funi}) is harmonic in
the nonlinear strain $u_{ij}$. Higher order terms in $u_{ij}$ are, of
course, permitted and are in fact necessary to preserve full
rotational invariance in $S_R$, which is present (but hidden), if the
uniaxial asymmetry arises as a result of the {\em spontaneous}
symmetry breaking of an {\em isotropic} state, as happens in nematic
elastomers, introduced in Sec.\ \ref{strainonly}.

In semi-soft elastomers\cite{FinKun97,WarTer96}, the rotational
invariance of the soft-elastomer isotropic state (discussed next)
is only weakly broken. Any model describing these systems must
introduce anisotropy in such a way that both the isotropic and the
anisotropic soft phases are reproduced when the anisotropy is set
to zero.  The simplest such model can be constructed by adding an
anisotropic term
\begin{equation}
f_{\rm anis} = - h n_{0i} {\tilde u}_{ij} n_{0j},
\end{equation}
which breaks $O_R$ symmetry, to the free energy of Eq.\
(\ref{eq:f}). Here ${\bf n}_0$ is a vector in $S_R$ that specifies
the direction of preferred alignment, and $h$ is a field measuring
the anisotropy strength.  The properties of this model will be
explored in a separate publication\cite{semi-soft}.

\section{Strain-only Model of Nematic Elastomers}
\label{strainonly}
Under appropriate conditions, for example for sufficiently small shear
modulus $\mu$ in the model free energy of Eq.(\ref{f_1}), there can be
a transition from an isotropic state with $\mm{\Lambda} \sim
\mm{\delta}$ to a uniaxial one with two rather than one distinct
eigenvalues for $\mm{\Lambda}$.  This nematic-gel state is obtained
from the isotropic one by stretching or compressing along some
arbitrary direction in $S_R$ specified by a unit vector $\nv_0$, which
without loss of generality we take to be along the $z$ axis.  It is
characterized by an anisotropic equilibrium right strain tensor
$\mm{u}_0$ with principal axis along $\nv_0$.  The transition to the
nematic gel can thus be described completely in terms of the free
energy $f(\mm{u} )$.  Alternatively, the nematic gel can be
characterized by an anisotropic equilibrium left strain tensor
$\mm{v}_0$ with anisotropy axis along some unit vector $\nv_1$ is
$S_T$, and the transition to it can be described by $f(\mm{v})$.  The
nematic gel breaks {\em both} $O_R$ and $O_T$ symmetry. The
description in terms of $\mm{u}$ displays explicitly the broken $O_R$
symmetry and that in terms of $\mm{v}$ the broken $O_T$ symmetry of
the nematic gel.  The underlying order parameter, however, is the
deformation tensor $\mm{\Lambda}$, which exhibits both broken $O_R$
and $O_T$ symmetry in the nematic gel.  Even though the nematic gel
breaks two symmetries, they are both broken at the same time, and
there is only one transition from the isotropic phase to the nematic
gel.  As discussed in Sec.\ \ref{iso-strain}, $f(\mm{u})$ and
$f(\mm{v})$ are identical functions of their arguments, and $\mm{u}$
and $\mm{v}$ develop nonzero anisotropic values simultaneously.

Though free energies expressed in terms of the strain $\mm{u}$ and
$\mm{v}$ provide complete descriptions of the phase transition to the
nematic gel, it is important to remember that the full position
function $\Rv( \xv)$ or equivalently the displacement $\uv(\xv)$ is
needed to describe all configurations of the gel.  The tensors
$\mm{\Lambda}$, $\mm{u}$, and $\mm{v}$ only provide information about
long-wavelength distortions.  A full statistical mechanical treatment
of nematic gels requires the inclusion of curvature energies depending
on the second derivative of $\Rv(\xv)$ into the elastic energy that
appears in the partition function trace. This will be discussed more
detail in a separate publication\cite{membraneXing}.

In this section, we will explore the properties of the spontaneously
formed nematic gel described in terms of $\mm{u}$ and $f(\mm{u})$.
The description in terms of $\mm{v}$ is essentially equivalent. We
will explicitly derive the soft elasticity of nematic gels whereby the
strain elastic constant $C_5$ [Eq.\ (\ref{funi})] vanishes
identically\cite{GolLub89} and there is zero stress\cite{VerWar95}
associated with appropriate strains up to a critical value
perpendicular and parallel to $\nv_0$ as long as other
strains\cite{Mitchell,dirrelax} are allowed to relax to their lowest
energy configurations. Our treatment provides a complete description
of nematic gels and transitions to them without any reference to
undelying nematic order.  In the next section, we will consider
nematic order and its coupling to strain and show that instabilities
toward the development of nematic order drive the decrease in the
shear modulus discussed in the preceding section.

\subsection{Description in terms of $u_{ij}$}

It is quite clear from the cubic form of the elastic free energy,
Eqs.\ (\ref{f_1}) and (\ref{eq:f_2}), that when $\mu$ becomes
sufficiently small, for finite $C$, there is a first-order
transition from an isotropic to a uniaxially distorted elastic
state, which is very similar to the familiar isotropic-to-nematic
transition\cite{deGennesProst93,Chandrasekhar92}.  We will
consider this transition in more detail in Sec.\
\ref{iso-uni-tran}.  In this subsection, we will investigate the
resulting anisotropic elastic state, whose properties depend only
on the existence of {\em spontaneously} formed anisotropy and not
on any particular model of the isotropic-to-nematic transition.

In the positive (negative) uniaxial state that results from such
transition, the elastic material is stretched (compressed) along
$\nv_0$ in $S_R$ and compressed (stretched) along directions
perpendicular to $\nv_0$.  This anisotropy axis can point in any
direction $\nv_1$ in $S_T$. For the moment, we will take $\nv_1 =
\nv_0$, i.e., we do not rotate the sample after it has been
stretched. In this case, the coordinates of its mass points in
$S_T$ are, therefore, described by
\begin{equation}
\Rv_0 (\xv )\equiv \xv' = \xv + \uv_0 = \mm{\Lambda}_0 \xv ,
\end{equation}
where the deformation tensor is spatially uniform and given by
\begin{mathletters}
\begin{eqnarray}
\Lambda_{0ij}&=& \Lambda_{0\perp}\delta^{0\perp}_{ij} + \Lambda_{0||}n_{0i} n_{0j},\\
&=&\Lambda_{0\perp} \delta_{ij} + (\Lambda_{0||}
- \Lambda_{0\perp})n_{0i} n_{0j},
\label{eq:Lam0}
\end{eqnarray}
\end{mathletters}
where $\delta^{0\perp}_{ij}=\delta_{ij}-n_{0i} n_{0j}$.  The
corresponding right equilibrium strain tensor is
\begin{equation}
\mm{u}_0 = \case{1}{2} (\mm{\Lambda}_0^T\mm{\Lambda}_0 -
\mm{\delta}) .
\label{eq:uo}
\end{equation}
The anisotropy of the uniaxial state can be characterized by the
anisotropy ratio\cite{warnerr}
\begin{equation}
r = {\Lambda_{0||}^2 \over \Lambda_{0\perp}^2} .
\label{eq:a-ratio}
\end{equation}
Since
\begin{equation}
\mm{\Lambda}_0^T\mm{\Lambda}_0 = \Lambda_{0\perp}^2[\mm{\delta} +
(r-1) {\bf n}_0 {\bf n}_0],
\end{equation}
it is clear that the system is isotropic if $r=1$ and only
anisotropic if $r\neq 1$.  Both positive ($r>1$) and negative
($r<1$) uniaxial anisotropies are possible, but we will focus
mostly on positive uniaxial systems.  In incompressible systems,
$\det \mm{\Lambda}_0 = \Lambda_{0||}\Lambda_{0\perp}^2 = 1$,
$\Lambda_{0\perp}=\Lambda_{0||}^{-1/2}$, and $r =
\Lambda_{0||}^3$.  Many of the properties of the uniaxial phase
depend critically on $r$.

The goal of this subsection is to explore the elasticity of this
spontaneously uniaxially-ordered gel.  Distortions such a system
can be described by deviations
\begin{equation}
\delta \mm{u} = \mm{u} - \mm{u}_0 = \case{1}{2}(\mm{\Lambda}^T
\mm{\Lambda} - \mm{\Lambda}_0\mm{\Lambda}_0 )
\label{eq:du}
\end{equation}
of the strain tensor $\mm{u}$ from its new equilibrium value
$\mm{u}_0$, both measured in the coordinates $\xv$ of the original
isotropic state, $S_R$.  It is, however, more common and
convenient to describe these distortions in terms of displacements
$\Rv'(\xv')\equiv\Rv(\xv)$ and strains $\mm{u}'(\xv')$ expressed
as functions of the coordinates $\xv'\equiv\Rv_0(\xv)$ of the new
equilibrium stretched state:
\begin{mathletters}
\begin{eqnarray}
\Rv'(\xv') & = & \xv' + \uv' ( \xv' ) = \xv + {\bf u}_0 +
\delta {\bf u} ( \xv )\label{eq:R'},\\
\mm{u}'& = &\case{1}{2}(\mm{\Lambda}^{\prime T}
\mm{\Lambda}^{\prime} - \mm{\delta})\approx
\case{1}{2}(\mm{\eta}^{\prime} + \mm{\eta}^{\prime T} ) ,
\label{eq:u'}
\end{eqnarray}
\end{mathletters}
where $\Lambda_{ij}' = \partial R'_i /\partial x'_j$ and $\eta'_{ij} =
\partial u'_i/\partial x'_j$.
Since
\begin{equation}
\Lambda_{ij} = {\partial R_i \over \partial x_j} = {\partial
R'_i \over \partial x'_k} {\partial x'_k\over \partial x_j} =
\Lambda'_{ik} \Lambda_{0kj} ,
\label{eq:lambda'}
\end{equation}
the strain deviation $\delta \mm{u}$ is directly proportional to
$\mm{u}'$ and therefore proportional to the symmetrized strain
$(\mm{\eta}' + \mm{\eta}^{\prime T})/2$ when linearized:
\begin{equation}
\delta \mm{u} = \mm{\Lambda}_0^T \mm{u}' \mm{\Lambda}_0 .
\label{eq:du-u'}
\end{equation}
One would normally expect the elastic free energy for strains
$\mm{u}'$ about the new uniaxial state to have the form of Eq.\
(\ref{funi}), characterized by five independent elastic
constants. However, as discussed in the Introduction, the fact
that the uniaxial state arose via {\em spontaneous} symmetry
breaking of an isotropic state guarantees that the shear modulus
$C_5$ {\em must} vanish.  We now demonstrate this explicitly.

Since the original free energy is invariant under rotations $\mm{O}_R$
in $S_R$, the anisotropy direction $\nv_0$ in $S_R$ is arbitrary, and
states characterized by strain $\mm{O}_R \mm{u}_0\ \mm{O}_R^{-1}$ and
$\mm{u}_0$ must have the same bulk energy.  This means that there is
no bulk energy cost associated with a strain
\begin{equation}
\mm{u}'(\theta) = (\mm{\Lambda}_0^T)^{-1} (\mm{O}_R \mm{u_0}\
\mm{O}_R^{-1} - \mm{u}_0 ) \mm{\Lambda}_0^{-1}
\label{eq:rot-inv}
\end{equation}
relative to the uniaxial state characterized by $\mm{u}_0$ since it
describes a rotation in $S_R$, and is therefore a Goldstone mode of
broken $O_R$ symmetry. For rotations through $\theta$ about the
$y$-axis,
\begin{equation}
\mm{O}_R = \left(
\begin{array}{ccc}
\cos \theta & 0 &\sin\theta \\
0 & 1 & 0 \\
-\sin \theta & 0 & \cos \theta
\end{array}
\right) .
\label{eq:rot-matrix}
\end{equation}
Using this $\mm{O}_R$ inside Eq.(\ref{eq:rot-inv}), we find that
$u_{ij}'(\theta)$ is a nontrivial strain even though it describes
a pure rotation in $S_R$.  Under this rotation, $u_{ij}'(\theta)$
only has components in the $xz$-plane (the plane of rotation
$\theta$) that are
\begin{mathletters}
\begin{eqnarray}
\mm{u'} &= &{1 \over 4} (r - 1) \left(
\begin{array}{cc}
1 - \cos 2\theta & r^{-1/2} \sin 2 \theta \\
r^{-1/2} \sin 2 \theta & - r^{-1} (1 - \cos 2 \theta )
\end{array}
\right) \label{eq:rot-u} \\
& \approx & {(r-1)\over 2\sqrt{r}} \left(
\begin{array}{cc}
0 & \theta \\
\theta & 0
\end{array}
\right) ,
\label{eq:rot-u_theta}
\end{eqnarray}
\end{mathletters}
where $r$ is the anisotropy ratio introduced in Eq.\
(\ref{eq:a-ratio}) and the final form is valid for small $\theta$.
Since, as just argued, the elastic free energy must be invariant
under rotations in $S_R$, it cannot depend on the rotation angle
$\theta$, and, therefore, there must be no energy cost associated
with an infinitesimal strain $u_{xz}' = u_{zx}'$.  Similarly,
invariance with respect to rotations about the $x$ axis implies no
energy cost associated with the strain $u_{yz}'$. Thus, the shear
elastic modulus $C_5$ must identically vanish in a {\em
spontaneously} uniaxial state, whose {\em harmonic} elastic
energy,
\begin{eqnarray}
f_{\rm uni}^N &= &\case{1}{2} C_1 u_{zz}^{\prime 2} + C_2 u'_{zz}
(u'_{xx} + u'_{yy} ) +
\case{1}{2}C_3 (u'_{xx} + u'_{yy})^2 \nonumber \\
&+& C_4 (u_{xx}^{\prime 2} + u_{yy}^{\prime 2} + 2 u_{xy}^{\prime
2}) ,
\label{funi2}
\end{eqnarray}
is characterized by only four elastic constants. The superscript
$N$ in $f_{\rm uni}^N$ is introduced to distinguish it from the
standard uniaxial energy $f_{\rm uni}$ [Eq.(\ref{funi})] with $C_5
\neq 0$. Because $f_{\rm uni}^N$ contains only quadratic terms in
strain $u_{ij}'$ relative to the broken-symmetry uniaxial phase,
as in similar systems\cite{comment_invariance} it is only
invariant with respect to {\em infinitesimal} rotations in $S_R$
and terms nonlinear in $u_{ij}'$ must be incorporated in order to
encode the full $O_R$ invariance\onlinecite{unpublished}.

There are striking experimental consequences of the existence of
zero-energy strains $\mm{u}'(\theta)$ given by Eqs.\
(\ref{eq:rot-inv}), (\ref{eq:rot-u}) for arbitrary $\theta$.
Namely, if one of the components of strain $u'_{xz}$, $u'_{xx}$,
or $u'_{zz}$ is imposed with the right sign, the other two
components can (boundary conditions of the experiment permitting)
adopt values to produce the zero-energy rotational strain of Eq.\
(\ref{eq:rot-u}).  When $r>1$, this relaxation is possible only
for positive $u'_{xx}$ (extension perpendicular to the uniaxial
direction ${\bf n}_0$), negative $u'_{zz}$ (compression along
${\bf n}_0$), and either positive or negative $u'_{xz}$.  For
negative anisotropy systems ($r<1$), the zero-energy strain
relaxation is possible only for negative $u'_{xx}$ and positive
$u'_{zz}$.

To illustrate this, consider first a sample with ${\bf n}_0$
aligned along the $z$-axis, with $r>1$.  From Eq.\
(\ref{eq:rot-u}), it follows that $u'_{xx}= (1/2)(r-1) \sin^2
\theta$ is positive for a rotational strain when $r>1$.  Thus, we
can only have soft elasticity for extensional strains along $x$,
and we take $u'_{xx}>0$. If no relaxation of strain is allowed,
this stretch would cost an energy proportional to $u_{xx}^{\prime
2}$.  If, however, strain relaxation is allowed, strains
\begin{eqnarray}
u_{zz}' & = & - {1 \over r} u_{xx}' \nonumber \\
u_{xz}' & = & \pm {1\over \sqrt{2r}} \sqrt{u'_{xx}(r-1 - 2
u_{xx}')}
\end{eqnarray}
convert the $u_{xx}'$ strain to a zero-energy rotation strain tensor
(Goldstone mode) with rotation angle
\begin{equation}
\theta = \sin^{-1} \sqrt{2 u'_{xx}\over r-1}.
\label{eq:theta}
\end{equation}
Thus, in an ideal system, there is no bulk energy cost associated with
strains $0 < u_{xx}' <(r-1)/2$.

\begin{figure}
\centerline{\epsfbox{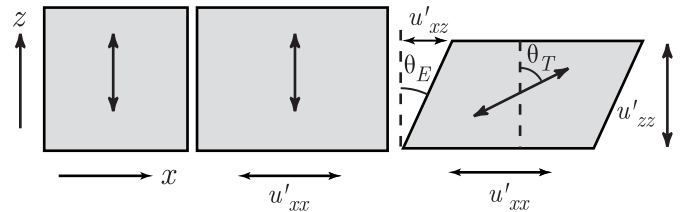}} \caption{Schematic
representation of the transformations of a rectangular piece of
soft nematic gel subjected to a strain perpendicular to its
anisotropy axis. If $u'_{zz}$ and $u'_{xz}$ are constrained to be
zero, there is no rotation of the anisotropy axis indicated by a
double arrow as shown in the middle figure. If these quantities
are allowed to relax, the anisotropy axis rotates to produce a
state with the same energy as the initial state as shown in the
final figure. In the process, the original rectangle is
transformed into a parallelogram sheared through and angle
$\theta_E$, and its anisotropy axis represented by the double
arrow is rotated through an angle $\theta_T$.}
\label{fig:rotate}
\end{figure}

\begin{figure}
\centerline{\epsfbox{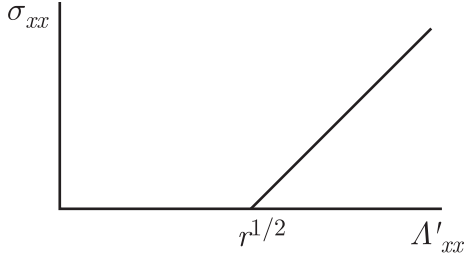}}
\caption{Stress $\sigma_{xx}$ versus strain $\Lambda_{xx}$ for an
  ideal soft nematic elastomer stretched along a direction
  perpendicular to the direction of initial alignment. The stress is
  zero up to a critical strain $\Lambda_{xx}= \sqrt{r}$.  Beyond that,
  the stress initially grows linearly from zero.}
\label{fig:stress-strain}
\end{figure}

The angle $\theta$ specifies the direction of the induced uniaxial
equilibrium stretch axis relative to the $z$-axis of a fixed
coordinate system in $S_R$.  In the current problem, this
anisotropy axis is initially along $z$, and it rotates toward the
$x$-axis as $u'_{xx}$ is increased until, at the critical strain
$u'_{xx} = (r-1)/2$, $\theta = \pi/2$ and the anisotropy axis has
been rotated to be along the $x$-axis, as illustrated in Fig.
\ref{fig:rotate}. For strains $u'_{xx}$ larger that $(r-1)/2$,
the sample will merely stretch along its new anisotropy axis
along $x$ with the additional strain $\delta u'_{xx} = u'_{xx} -
(r-1)/2$.  We can calculate the energy associated with this
additional strain from the harmonic free energy of Eq.\
(\ref{funi2}) provided we remember that $\delta u'_{xx}$ is
measured relative to the original reference system with the
anisotropy axis along $z$ rather than $x$, i.e., if we remember
to replace $u'_{zz}$ in Eq.\ (\ref{funi2}) by ${\overline
u}'_{zz} = (\Lambda_{0\perp}^2 / \Lambda_{0||}^2)\delta u'_{xx} =
\delta u'_{xx} /r$.  $u'_{xx}$ and $u'_{yy}$ should be rescaled
as well, but since we minimize over these quantities at fixed
${\overline u}'_{zz}$, we do not have to explicitly consider these
rescalings.  Performing this minimization, we find
\begin{equation}
\delta f = \cases{0,&if $\delta u'_{xx}<0$;\cr {1 \over 2 r^2}
\left( C_1 - {C_2^2\over 2C_4+ 2C_3} \right) (\delta u'_{xx})^2,
&if $\delta u'_{xx}>0$.\cr}
\end{equation}
Consequently, the stress $\partial f/\partial \Lambda'_{xx} =
\Lambda'_{xx} \partial f/\partial u'_{xx} $ for an ideal nematic
gel is zero for $u'_{xx} < (r-1)/2$ and grows linearly in $\delta
u'_{xx}$ for $u'_{xx} > (r-1)/2$ as shown in Fig.\
\ref{fig:stress-strain}.

The angle $\theta_T$ of rotation of the anisotropy axis in the
target space $S_T$ is not the same as the rotation angle $\theta$
in $S_R$.  Indeed, since the energy is invariant with respect to
rotations in $S_T$, $\theta_T$ would be arbitrary, were it not for
boundary conditions. A specific experimental geometry might for
example demand that there be no change in the $z$ coordinates of
mass points as a function of $x'$.  This is the situation depicted
in Fig.\ \ref{fig:rotate}. In this case, $\Lambda'_{zx}=0$, but
$\Lambda'_{xz} \neq 0$, and the $xz$ submatrix of $\Lambda'_{ij}$
takes the form
\begin{equation}
\mm{\Lambda}' = \left(
\begin{array}{cc}
\Lambda'_{xx} & \Lambda'_{xz} \\
0 & \Lambda'_{zz}
\end{array}
\right).
\label{Lambda_prime}
\end{equation}
This deformation tensor corresponds to a zero-energy rotation in
$S_R$ described by a rotation matrix $\mm{O}_R$ [Eq.\
(\ref{eq:rot-matrix})] provided there exists a rotation matrix
\begin{equation}
\mm{O}_T = \left(
\begin{array}{cc}
\cos \theta_T & \sin \theta_T \\
- \sin\theta_T & \cos \theta_T
\end{array}
\right)
\end{equation}
in the $xz$-plane of $S_T$ such that the strain
$\mm{\Lambda}=\mm{\Lambda}' \mm{\Lambda}_0$ relative to the
original reference system is
$\mm{O}_T\mm{\Lambda}_0\mm{O}_R^{-1}$, or
\begin{equation}
\mm{\Lambda}' = \mm{O}_T\mm{\Lambda}_0\mm{O}_R^{-1}\mm{\Lambda}_0^{-1},
\end{equation}
an expression that is fully consistent with the form of strain
Goldstone mode $\mm{u}'(\theta)$, Eq.\ \ref{eq:rot-u}.  A
straightforward calculation yields
\begin{equation}
\Lambda'_{zx} =  - \cos \theta \sin \theta_T +\sqrt{r} \sin \theta
\cos \theta_T,
\end{equation}
which, upon imposition of the boundary condition of $\Lambda'_{zx}=0$,
Eq.(\ref{Lambda_prime}), gives
\begin{mathletters}
\begin{eqnarray}
\sin^2 \theta_T &=& {r \sin^2 \theta \over 1 + (r-1) \sin^2 \theta }\\
&=& {r\over r-1} \left( 1 - {1 \over \Lambda_{xx}^{\prime 2}}\right) ,
\label{eq:sinthetaT}
\end{eqnarray}
\end{mathletters}
where we used $\sin^2 \theta=2u_{xx}'/(r-1)$, Eq.\ (\ref{eq:theta})
and $(\Lambda_{xx}')^2 - 1 = 2 u_{xx}'$ to obtain the final form. This
is exactly the same result obtained via a direct minimization of the
neoclassical rubber energy in the incompressible limit\cite{VerWar96}.

The angle $\theta_T$ is simply the angle that the uniaxial
anisotropy axis makes with the $z$ axis in the target space.  This
axis is the principal axis of $\mm{\Lambda} \mm{\Lambda}^T$ (or
equivalently of $\mm{v}$).  A direct calculation of $\mm{\Lambda}'
\mm{\Lambda}_0 \mm{\Lambda}_0^T \mm{\Lambda}^{\prime T}$ yields
\begin{equation}
\mm{\Lambda} \mm{\Lambda}^T = \Lambda_{0\perp}^2 \left(
\begin{array}{cc}
\cos^2 \theta_T + r \sin^2 \theta_T & (r-1) \sin \theta_T \cos\theta_T\\
(r-1) \sin\theta_T \cos\theta_T & r \cos^2 \theta_T + \sin^2
\theta_T ,
\end{array}
\right)
\end{equation}
which is nothing more than $\mm{\Lambda_0}\mm{\Lambda_0}^T$
rotated though $\theta_T$ so that the principal axis is along
$\nv_1 = (\sin \theta_T , \cos \theta_T)$.  Under the
transformation defined by $\mm{\Lambda}'$ of Eq.\
(\ref{Lambda_prime}), a rectangle will be transformed into a
parallelogram with two edges parallel to the $x$-axis and two
edges making an angle
\begin{equation}
\theta_E = \tan^{-1} {\Lambda_{xz}'\over \Lambda_{zz}'}
=\tan^{-1}\left( {(r -1) \tan \theta_T \over r + \tan^2
\theta_T}\right)
\end{equation}
with the vertical.  Both $\theta_E$ and $\theta_T$ are indicated
in Fig.\ \ref{fig:rotate}.  Note that the argument of the inverse
tangent in the expression for $\theta_E$ rises from zero, reaches
a maximum, and then returns to zero as $\theta_T$ passes from zero
to $\pi/2$, the maximum angle of rotation of the anisotropy axis.
In the process, $\mm{\Lambda}$ passes from having extension
$\Lambda_{0||}$ along the $z$-axis and $\Lambda_{0\perp}$ along
the $x$-axis to having extension $\Lambda_{0\perp}$ along the
$z$-axis and $\Lambda_{0||}$ along $x$.  Thus, the original
rectangle is distorted to a parallelogram that first becomes more
slanted as strain increases, reaches a maximum slant, and then
becomes less slanted until it finally reaches a rectangular form
that is precisely the original rectangle rotated by $\pi/2$ before
stretching further along the $x$ axis.

It is notable that the expression, Eq.\ (\ref{eq:sinthetaT}), for
$\sin \theta_T$ is independent of the detailed form of the elastic
energy and that it is characterized only by the level of
anisotropy $r$ of the initial uniaxial state and by the
deformation $\Lambda'_{xx}$ applied to it.  The advantage of our
approach is that it makes it clear that the phenomenon of soft
response, summarized by Fig.\ \ref{fig:stress-strain}, i.e., zero
stress for a range of longitudinal strain, applied perpendicular
to the uniaxial direction, follows entirely from general symmetry
principles of breaking of rotational invariance of the reference
state. It is a property of {\em any} nematic gel formed
spontaneously from an isotropic gel, and is therefore independent
of the details of the microscopic model of the gel and the
mechanism that drives the uniaxial instability, as confirmed by
the generic form for $\theta_T$ in Eq.\ (\ref{eq:sinthetaT}).

There is spectacular experimental evidence\cite{FinKun97} for
$\theta_T(\Lambda_{xx}')$, Eq.\ (\ref{eq:sinthetaT}). Likewise,
experiments confirm the GL prediction of softness (vanishing
stress up to a critical value of strain, Fig.\
\ref{fig:stress-strain}), in the limit in which the gel is
crosslinked in the isotropic phase.  In contrast, chemically
identical networks crosslinked in the nematic phase exhibit a
plateau in the stress-strain curve that approaches zero as the
degree of nematic order during crosslinking is decreased. See
Ref.\ \onlinecite{Warner99} for a presentation and discussion of
these results.

It is clear that for $r>1$ an imposed shear strain $u'_{xz}$ or a
compressional strain $u'_{zz}<0$ can be converted to zero-energy
rotational strain just as for the case $u'_{xx}>0$ just
considered. The energy of a shear strain is zero for
$|u'_{xz}|<(r-1)/(4\sqrt{r})$, and that of a compressional strain
is zero for $|u'_{zz}|< (r-1)/(2r)$. While softness with respect
to {\em small} shears below a critical frequency has been
observed\cite{dyn}, to our knowledge, our above prediction of
softness over a {\em finite} range of strain for the geometries
with imposed $u'_{xz}\neq 0$ or $u'_{zz}<0$ has not been tested.

The above discussion and the structure of the zero-energy strain
$u'_{ij}(\theta)$, Eq.(\ref{eq:rot-u}), imply that our arguments
for soft elasticity go through equally well for the negative
uniaxial anisotropy elastomers, $r<1$, but with reversed signs of
the imposed zero-energy strains. Thus, in negative uniaxial
elastomers, there is no stress associated with unconstrained {\em
compression} along $x$, $u'_{xx}<0$ and {\em extension} along $z$,
$u'_{zz}>0$.

\subsection{Description in terms of $v_{ij}$}

In the discussion just presented, the nematic state and its soft
elasticity were described in terms of the right strain tensor
$u_{ij}$.  As demonstrated in Sec.\ref{iso-strain}, this state
could equally well be described in terms of the left strain tensor
$v_{ij}$, which has an equilibrium value
\begin{equation}
\mm{v}_0 = \case{1}{2} (\mm{\Lambda}_0 \mm{\Lambda}_0^T -
\mm{\delta} )
\end{equation}
that is identical to $\mm{u}_0$ in the basis defined by $\nv_0$.
The deviations from the spontaneously anisotropic equilibrium
state $\mm{\Lambda}_0$ can be described by
\begin{equation}
\delta \mm{v} = \mm{v} - \mm{v}_0 = \case{1}{2}(\mm{\Lambda}
\mm{\Lambda}^T - \mm{\Lambda}_0 \mm{\Lambda}_0^T )
\label{eq:dvdef}
\end{equation}
as well as $\delta \mm{u}$.  The deviations $\delta\mm{u}$ and
$\delta\mm{v}$, however, have different relations to the displacement
gradient tensor $\mm{\eta}'$ relative to the anistropic equilibrium
state.  As we saw in Eq.\ (\ref{eq:du-u'}), $\delta\mm{u}$ is linearly
proportional to $\mm{u}'$ and at linear order depends only on the {\em
  symmetrized} part of $\mm{\eta}'$.  $\delta \mm{v}$ on the other
hand is {\em not} proportional to $\mm{v}'$ defined by Eq.\
(\ref{eq:strain2}) with $\mm{\Lambda}$ replaced by
$\mm{\Lambda}'$, and to linear order, it depends on {\em both} the
symmetric and anti-symmetric parts of $\mm{\eta}'$. Using Eqs.\
(\ref{eq:dvdef}) and (\ref{eq:Lam1}), we easily derive
\begin{eqnarray}
\Lambda_{0\perp}^{-2} \delta v_{ij} & = & \case{1}{2}( \eta_{ij}'
+
\eta_{ji}' + \eta_{ik}' \eta_{jk}' ) \nonumber \\
&  & + \case{1}{2} (r-1 ) (n_{0i} n_{0k} \eta_{jk}' + \eta_{ik}'
n_{0k} n_{0j} \nonumber \\
& & + \eta_{ik}' n_{0k} n_{0l} \eta_{jl}' ),
\label{eq:dveta}
\end{eqnarray}
in terms of $\mm{\eta}'$, with $r$ the previously defined
anisotropy ratio [Eq.\ (\ref{eq:a-ratio}].  Using the
decomposition
\begin{equation}
\mm{\eta}' = \mm{\eta}'_S + \mm{\eta}'_A \qquad
\mm{\eta}^{\prime T} = \mm{\eta}'_S - \mm{\eta}'_A ,
\label{eq:etaAS}
\end{equation}
where $\eta'_{Sij} = \eta'_{Sji}$ and $\eta'_{Aij} = -
\eta'_{Aji}$ are, respectively, the symmetric and anti-symmetric
parts of $\eta'_{ij}$ and defining a rotation angle
\begin{equation}
\Omega_i = \case{1}{2}\epsilon_{ijk} {\partial u'_k\over
\partial x'_j} = \case{1}{2} \epsilon_{ijk} \eta'_{Akj} ,
\label{eq:Omega}
\end{equation}
we obtain
\begin{eqnarray}
\Lambda_{0\perp}^{-2} \delta v_{ij} & = & \eta'_{Sij}
\nonumber \\
& & + \case{1}{2} (r-1)[n_{0i} n_{0k} \eta'_{Sjk} + \eta'_{Sik}
n_{0k} n_{0j}]
\label{eq:dv}\\
& & -\case{1}{2} (r-1) [ n_{0i} ({\mathbf n}_0 \times {\mathbf
\Omega})_j +({\mathbf n}_0 \times {\mathbf \Omega})_i
n_{0j}]\nonumber
\end{eqnarray}
to linear order in $\eta'_{ij}$.  We thus explicitly demonstrate
that $\delta\tu_{ij}$ is a function of both the symmetric and
anti-symmetric parts of $\eta'_{ij}$.  At first pass, this
observation appears to contradict the facts that the linearized
form of $\delta u_{ij}$, Eqs.\ (\ref{eq:u'}) and (\ref{eq:du-u'}),
does not depend on $\eta'_{Aij}$, and that the harmonic free
energy of the nematic phase must have exactly the same form
whether expressed in terms of $\delta u_{ij}$ or $\delta
\tu_{ij}$. The dilemma is resolved by noting that only the soft
components of the strain, $\delta \tu_{xz}$ and $\delta \tu_{yz}$
depend on $\eta'_{Aij}$, and that, to harmonic order, $f_{\rm
uni}^N (\mm{v})$ is guaranteed by $O_R$ invariance (which leads to
vanishing of $C_5$) to be independent of such strains.
Consequently, consistent with the expectations, in both $u_{ij}$
and $v_{ij}$ descriptions no anti-symmetric part of the
deformation tensor $\eta'_{Aij}$ appears.

\subsection{Isotropic-to-uniaxial transition}
\label{iso-uni-tran}

In a transition from the isotropic to the uniaxial state, the strain
develops a nonvanishing anisotropic component.  We can describe this
transition equivalently in terms of $\mm{u}$ and $f(\mm{u})$ or in
terms of $\mm{v}$ and $f(\mm{v})$.  To be concrete, we will use
$\mm{u}$-description here. Since the isotropic part of the strain is
insensitive to anisotropy, the appropriate order parameter for the
transition is the symmetric-traceless component of the strain,
$\mm{\ut}$ [Eq.\ (\ref{eq:st-strain})], which is identical in form to
the symmetric-traceless order-parameter tensor $Q_{ij}$ of a nematic
liquid-crystalline phase\cite{deGennesProst93,Chandrasekhar92}. Thus,
in mean-field theory, the transition from the isotropic to the
uniaxial state is identical to the isotropic-to-nematic transition,
whose properties have been exhaustively
studied\cite{deGennesProst93,Chandrasekhar92}. (In Appendix\
\ref{app:A}, we review the formal properties of this transition that
are relevant to the current discussion.) To see this in more detail,
we can integrate out the ``massive'' $\Tr \um$ from $f(\mm{u})$ in
Eq.\ (\ref{eq:f}), to obtain an effective theory in terms of $\ut$
alone.  This operation yields
\begin{equation}
\Tr \um = {E \over B} \Tr \mm{\ut}^2 ,
\end{equation}
and the effective free energy reduces to $f_1 (\ut )$ of Eq.\
(\ref{eq:f_2}). Because of the presence of a cubic invariant, the
free energy $f_1(\mm{\ut})$ exhibits a first-order transition at
$A = A_c=C^2/(12D)$ to a state with
\begin{equation}
\ut^0_{ij} = \psi (n_i^0 n_j^0 - \case{1}{3} \delta_{ij}) ,
\end{equation}
where $\psi$ satisfies the equation of state
\begin{equation}
A -C \psi +\case{8}{3}D \psi^2 = 0 .
\label{eqn_of_state}
\end{equation}
The total strain in the distorted state is thus
\begin{equation}
u^0_{ij}=\ut^0_{ij}+ {2 \over 9} {E\over B} \psi^2
\delta_{ij} .
\label{eq:u0}
\end{equation}
This corresponds to a stretched state with target space
positions $\Rv_0 = \Lm_0 \xv$ characterized by a deformation
tensor
\begin{equation}
\Lm_0 =  \sqrt{1 + 2 \um^0} =
\left(
\begin{array}{ccc}
    \Lambda_{0\perp} & 0 & 0 \\
    0 & \Lambda_{0\perp} & 0 \\
    0 & 0 & \Lambda_{0||}
\end{array}
\right) ,
\end{equation}
with
\begin{eqnarray}
\Lambda_{0 \perp} & = & \left(1+{4 \over 9}{E\over B} \psi^2
-{2\over 3} \psi\right)^{1/2} \nonumber \\
\Lambda_{0||}& = & \left(1+ {4\over 9}{E\over B} \psi^2 + {4
\over 3} \psi \right)^{1/2} .
\end{eqnarray}
This form for $\mm{\Lambda}_0$ preserves the volume up to order
$\psi^2$. The order parameter $\psi$ is a direct measure of the
spontaneous stretch anisotropy of the nematic state, with
\begin{equation}
\psi = \case{1}{2}(\Lambda_{0||}^2 - \Lambda_{0\perp}^2 ) =
\case{1}{2}\Lambda_{0\perp}^2 (r-1) ,
\end{equation}
where $r$ is defined in Eq.\ (\ref{eq:a-ratio}.

The elastic free energy $f_1(\mm{u})$ can now be expanded in
powers of $\delta \mm{u} = \mm{u} - \mm{u}_0$ and reexpressed in
terms of $\mm{u}'$, defined in Eq.\ (\ref{eq:du-u'}). Ward
identities, imposed by the rotational $O_R$ invariance, guarantee
that terms proportional to $(\delta u_{xz})^2$ and $(\delta
u_{yz})^2$ vanish. The harmonic elastic energy is, therefore,
given by $f_{\rm uni}^N$ [Eq.\ (\ref{funi2})] with
\begin{eqnarray}
C_1 & = & \left[B\left(1 - {4E \over 3B}\psi\right)^2 - {4 \over
3}
A + {2 \over 3} C \psi\right] \Lambda_{0||}^4 \nonumber \\
C_2 & = & \left[B\left(1 - {4 E\over 3B}\psi\right)\left(1 + {2
E\over 3B}\psi \right) + {2 \over 3} A - {1 \over 3} C
\psi\right] \Lambda_{0||}^2 \Lambda_{0\perp}^2 \nonumber \\
C_3 & = & \left[B\left(1 + {2E \over 3B}\psi\right)^2 - {1 \over
3} A - {4 \over 3} C \psi\right] \Lambda_{0\perp}^4 \nonumber \\
C_4 & = & {3\over 2}C \psi \Lambda_{0\perp}^4,
\label{Cs}
\end{eqnarray}
with $C_5=0$, as anticipated in our early discussion of the generic,
symmetry-dictated form of the elastic free energy. From Eq.\
(\ref{eqn_of_state}) we note that $\psi$ has the same sign as $C$,
ensuring that $C_4 \sim C \psi$ is always positive.

\subsection{Biaxial nematic}

It is clear from the form of $f_{\rm uni}^N(\mm{u}')$, Eq.
(\ref{funi2}) that if $C_4$ is driven negative\cite{comment_C4}
the uniaxial state becomes unstable to strains in the $xy$-plane
perpendicular to the established uniaxial order, i.e., the
uniaxial state becomes unstable relative to a biaxial state with
different equilibrium strains in all three directions. A biaxial
nematic gel is softer than a unaxial one\cite{warnerbiax}, and, as
we will show here, it has no nonvanishing shear modulus in three
dimensions. The order parameter for the unaxial-to-biaxial
transition is the two-dimensional symmetric-traceless tensor
obtained by projecting $\mm{\tilde{u}}$ onto the $xy$-plane. Since
there are no cubic invariants of a two-dimensional symmetric
traceless tensor, the transition from the uniaxial to the biaxial
state is generically a continuous transition in the $xy$
universality class.

The biaxial phase is characterized by a Cauchy deformation tensor
with three independent components:
\begin{equation}
\mm{\Lambda}_0 =  \left(
\begin{array}{ccc}
    \Lambda_{01} & 0 & 0 \\
    0 & \Lambda_{02} & 0 \\
    0 & 0 & \Lambda_{03}
\end{array}
\right) ,
\end{equation}
and the corresponding equilibrium strain tensor given by
\begin{equation}
\mm{u}_0 = \left(
\begin{array}{ccc}
    u_{01} & 0 & 0 \\
    0 & u_{02} & 0 \\
    0 & 0 & u_{03}
\end{array}
\right),
\label{eq:biau0}
\end{equation}
with $u_{0\alpha} = (\Lambda_{0\alpha}^2 -1)/2$, $\alpha = 1,2,3$.
The additional broken rotational symmetry of the biaxial relative
to the uniaxial phase causes more shear elastic moduli to vanish.
As in uniaxial gels, strains of the form of Eq.\
(\ref{eq:rot-inv}) [with $\mm{u}_0$ given by Eq.\
(\ref{eq:biau0})] that arise from arbitrary three-dimensional
rotations $O_R$ in $S_R$ cost no energy.  For simplicity, we
consider only the most general {\em infinitesimal} rotation
matrix, which can be expressed in terms of rotation angles
$\theta_x$, $\theta_y$, and $\theta_z$, respectively about the
$x$, $y$, and $z$ axes as
\begin{equation}
O_R = \left(
\begin{array}{ccc}
1 & - \theta_z & \theta_y \\
\theta_z & 1 & - \theta_x \\
-\theta_y & \theta_x & 1
\end{array}
\right) .
\label{eq:biarot}
\end{equation}
The symmetric zero-mode strain tensor $\mm{u}'$ calculated from this
$O_R$ and Eq.\ (\ref{eq:rot-inv}) has components that to linear order
in the infinitesimal angles of rotation are $u_{xx}' = u_{yy}' =
u_{zz}' = 0$ and
\begin{mathletters}
\begin{eqnarray}
u_{xy}' & = & \Lambda_{01}^{-1} \Lambda_{02}^{-1} (u_{01} -
u_{02})\theta_z \nonumber \\
u_{xz}' & = & -\Lambda_{01}^{-1} \Lambda_{03}^{-1} (u_{01} -
u_{03})\theta_y \nonumber \\
u_{yz}' & = & \Lambda_{02}^{-1} \Lambda_{03}^{-1} (u_{02} -
u_{03})\theta_x  .
\end{eqnarray}
\end{mathletters}
Since the underlying $O_R$ invariance demands that there can be no
energy cost associated with such zero-mode strains, the elastic energy
cannot depend on the shear strains $u_{xy}'$, $u_{xz}'$, or $u_{yz}'$
to harmonic order. The hallmark property of solid that it can support
a static shear stress is therefore lost in a spontaneously biaxial
solid.  The biaxial nematic is an anisotropic tethered
fluid\cite{NelsonPeliti,AL,LR}. The harmonic elastic energy of a
biaxial gel, therefore, depends only on the compression/extensional
strains $u_{xx}'$, $u_{yy}'$ and $u_{zz}'$ and has the form
\begin{equation}
f_{\rm biax} = \case{1}{2}\sum B_{\alpha\beta} u_{\alpha \alpha}'
u_{\beta \beta}' .
\end{equation}
There are in general six independent components of $B_{\alpha
\beta}$.  As in uniaxial gels, there is soft compressional and
extensional elasticity in  biaxial gels with vanishing stress up
to critical values of the strain.  We will not treat these
properties in detail here.

\section{Nematic Gels: Strain and Orientational Order}
\label{nemgels}

In this section we extend our formulation of the model of
anisotropic gels to include both the elastic and orientational
(nematic) degrees of freedom.  We first consider a ``soft-spin"
theory in which orientational order is described by
symmetric-traceless nematic order parameter $Q_{ij}$, which has
both uniaxial and biaxial components. This theory, which can
describe both isotropic and anisotropic phases of gels and the
transitions between them, takes explicit account of the coupling
between strain and $Q_{ij}$. It can be viewed as a theory in which
the familiar isotropic-nematic transition characterized by
ordering of $Q_{ij}$ induces elastic distortion.  Guided by the
underlying rotational symmetry of the nematic gel, we then develop
a complementary ``hard-spin'' model of nematic gels valid deep in
the nematically ordered phase. This theory is formulated in terms
of the strain and the nematic director $\nv$ alone, with all
``massive'' modes (e.g., magnitude of the uniaxial order $S$ and
biaxial fluctuations) integrated out. A common feature of these
complementary models is their invariance with respect to global
simultaneous rotations of strain and nematic order. This
invariance leads to gauge-like couplings between strain and
nematic order, whose harmonic limit reduce to those derived by
Olmsted\cite{Olmsted94} following de Gennes\cite{deGennes}.
However, our expression in terms of $v_{ij}$ and $\nv$ for the
{\em globally} invariant energy deep in the nematic phase is new.

\subsection{Simple model of the $IN$ transition}

In the preceding section, we investigated a model in which an
isotropic elastic medium undergoes a spontaneous anisotropic
distortion triggered by the fall of its shear modulus below a
critical value.  In liquid-crystal elastomers, the reduction in
the shear modulus and the elastic distortion that it leads to are
actually driven by the underlying isotropic-nematic transition of
the mesogenic component of the gel, that is orientational ordering
of e.g., side-chain or main-chain nematogens.  It is, therefore,
of some interest to develop a model in which the orientational
order parameter $Q_{ij}$ explicitly appears.

A generic model free energy density for such a model of a
liquid-crystal gel will consist of an isotropic elastic term $f_{\rm
  el}(\um)$, a term $f'_Q(\mm{Q})$ for nematic orientational order,
and a nemato-elastic term $f_C(\vm,\mm{Q})$ that couples strain to the
nematic order parameter $Q_{ij}$:
\begin{equation}
f_{{\rm el}-Q} = f_{\rm el} + f'_Q + f_C .
\label{eq:fel-Q}
\end{equation}
For simplicity, we can take $f_{\rm el}$ to be the elastic energy
$f$ of Eq.\ (\ref{f_1}) with only quadratic-order terms in
$u_{ij}$ (or, equivalently, $\tu_{ij}$), and near the IN
transition, we can choose the usual Landau-de-Gennes form for
$f'_Q$:
\begin{equation}
f'_Q  = \case{1}{2} r'_Q \Tr \mm{Q}^2 - w_3 \Tr \mm{Q}^3 + w'_4
(\Tr \mm{Q}^2)^2 .
\label{eq:fQ'}
\end{equation}
Terms in gradients of $\mm{Q}$ should also be included, but they
do not affect the present mean-field discussion, and we will
therefore ignore them here.  The most general local energy
coupling strain to $Q_{ij}$ can be constructed from products of
terms invariant under arbitrary rotations in both $\SR$ and
$\ST$, whose general form is $\Tr[\mm{\tu}^{n_1} \mm{Q}^{m_1}
...\mm{\tu}^{n_p} \mm{Q}^{m_p}]$. Note that these terms involve
couplings between $\tu_{ij}$ (rather than $u_{ij}$) and $Q_{ij}$,
because $Q_{ij}$ exists in $\ST$, and like $\tu_{ij}$ transforms
like a tensor under rotations in $\ST$ but like a scalar under
$O_R$ rotations in $\SR$.  To keep our discussion simple, we will
for the moment consider a simple form for $f_C$:
\begin{equation}
f_C = - s \Tr \um \Tr \mm{Q}^2 - 2 t \Tr \mm{\ttu} \mm{Q} ,
\label{eq:fC}
\end{equation}
where $\ttu_{ij} = \tu_{ij} - \case{1}{3} \delta_{ij} \tu_{kk}$
is the symmetric-traceless part of $\tu_{ij}$ and where we used
the fact that $\Tr \um = \Tr\mm{\tu}$. This energy captures the
important qualitative features of strain-orientational coupling,
namely that the development of orientational order will drive an
anisotropic distortion and a smaller change in volume.

An elastic energy $f_{\rm el}^u$ that is a function of strain alone
can be obtained by integrating $\mm{Q}$ out of the total free energy
of Eq.\ (\ref{eq:fel-Q}).  The leading order correction of this
operation to $f_{\rm el}$ is $-2(t^2/r_Q'){\rm Tr}{\tilde u}^2$.  Thus
$f_{\rm el}^u$ has exactly the same form as Eq. (\ref{f_1}), with
$\mu$ replaced by $\mu'=\mu - (2 t^2/r_Q')$. Clearly $\mu'$ decreases
and passes through zero as $r_Q'$ decreases and the $IN$ transition is
approached from the isotropic phase. Thus, the decrease in $\mu$ in
the models of Sec.\ \ref{strainonly} arises from instabilities toward
the development of nematic orientational order.

To treat the effects of strain-orientational coupling after the
transition to the nematic state occurs, it is useful to recast $f$ in
a slightly different form:
\begin{eqnarray}
f_{{\rm el}-Q} & = & \case{1}{2} B[\Tr \um -(s/B)\Tr \mm{Q}^2]^2
\nonumber\\
&& + \mu \Tr[\mm{\ttu} - (t/\mu)\mm{Q}]^2 + f_Q
\label{eq:fel-Q2}
\end{eqnarray}
where
\begin{equation}
f_Q = \case{1}{2} r_Q \Tr \mm{Q}^2 - w_3 \Tr \mm{Q}^3 + w_4 (\Tr
\mm{Q}^2)^2
\label{eq:fQ}
\end{equation}
with $r_Q = r'_Q - 2(t^2/\mu)$ and $w_4 = w'_4-(s^2/2B)$. This
free energy leads to the equations of state
\begin{eqnarray}
{\partial f\over \partial \Tr\mm{u}} & = & B[\Tr \mm{u} -
(s/B)\Tr\mm{Q}^2 ]= 0\nonumber \\
{\partial f \over \partial \ttu_{ij}} & = & \mu[\ttu_{ij} -
(t/\mu) Q_{ij}] = 0\\
{\partial f \over \partial Q_{ij} } & = & {\partial f_Q \over
\partial Q_{ij}} - s[\Tr \mm{u} - (s/B)\Tr \mm{Q}^2] Q_{ij}
\nonumber \\
& & - 2t[\ttu_{ij} - (t/\mu) Q_{ij}]=0 .
\end{eqnarray}
The uniaxial solutions to these equations are given by:
\begin{mathletters}
\begin{eqnarray}
{\rm Tr\um_0 }&=& \frac{s}{B} \Tr\mm{Q}_0^2,\label{eq_a}\\
\ttu^0_{ij} &=& \frac{t}{\mu}Q^0_{ij},\label{eq_b}\\
Q^0_{ij} &=& S(n_{0i} n_{0j} - \case{1}{3}
\delta_{ij}),\label{eq_c}
\end{eqnarray}
\label{eq_solutions}
\end{mathletters}
with $S$ satisfying
\begin{equation}
r_Q S - w_3 S^2 +\case{8}{3} w_4 S^3 = 0.
\end{equation}
Using these uniaxial solutions Eq.\ (\ref{eq_solutions}) in the
new stretched state, we find
\begin{eqnarray}
\Lambda_{0\perp}^2 &= &1 + \frac{4}{9}\frac{s}{B} S^2 -
\frac{2}{3}\frac{t}{\mu} S
\nonumber\\
\Lambda_{0||}^2 & = & 1 + \frac{4}{9}\frac{s}{B} S^2 +
\frac{4}{3}\frac{t}{\mu} S .
\label{Lambda2}
\end{eqnarray}
Note that $\Lambda_{0||}^2 - \Lambda_{0\perp}^2 = 2 (t/\mu) S$ is
linear in the nematic order parameter $S$.

As discussed in Appendix \ref{app:A}, fluctuations away from the
equilibrium state are conveniently treated with the introduction
of a complete set of five orthonormal symmetric-traceless matrices
$I^{\alpha}_{ij}$ satisfying $I^{\alpha}_{ij} I^{\beta}_{ji} =
\delta^{\alpha\beta}$ that allow us to expand $Q_{ij}$ and
${\tilde v}_{ij}$ as $Q_{ij} = \sum_{\alpha=0}^4 Q_{\alpha}
I^{\alpha}_{ij}$ and ${\tilde v}_{ij} = \sum_{\alpha=0}^4
\tu_{\alpha}I^{\alpha}_{ij}$. Expressions for $Q_{\alpha}$ and
$\tu_{\alpha}$ in terms of $Q_{ij}$ and $\tu_{ij}$, respectively
are given in Eqs.\ (\ref{Qalpha}). In particular, $Q_0 =
\sqrt{2/3} S$. In terms of these variables, we have to harmonic
order in $\delta \tu_{ij}$ and $\delta Q_{ij}$
\begin{eqnarray}
\delta f_{{\rm el}-Q} & = & \case{1}{2}B [\Tr \delta \um -(4 s
/3B) S \delta
S]^2 \nonumber \\
&& + \mu[\delta \tu_0 - (t/\mu)\sqrt{2/3} \delta S]^2 \nonumber \\
 & & + \mu\sum_{\alpha=1}^4 [\tu_{\alpha} -
(t/\mu) Q_{\alpha}]^2 \nonumber\\
& & + \case{1}{2} A_1 (\delta S)^2 + \case{1}{2} A_2[Q_1^2 +
Q_2^2 ] ,
\label{dfel-Q}
\end{eqnarray}
where $A_1$ and $A_2$ are given in Eqs.\ (\ref{AdfQ}). Rotational
invariance of $f_Q$ guarantees that terms $Q_3^2 \sim Q_{xz}^2$ and
$Q_4^2 \sim Q_{yz}^2$ do not appear in the nematic state. We can
integrate out the ``massive" longitudinal mode $\delta S$ and biaxial
modes $Q_1$ and $Q_2$ to obtain
\begin{eqnarray}
\delta f_v &= &\case{1}{2}{\overline B}_1 (\delta \tu_{zz})^2
+{\overline B}_2 \delta \tu_{zz} (\delta \tu_{xx} + \delta
\tu_{yy})\nonumber \\
& & + \case{1}{2}{\overline B}_3(\delta \tu_{xx} + \delta
\tu_{yy})^2  \label{eq:fv1}\\
& & + {\overline B}_4 (\delta \tu_{xx}^2 + \delta \tu_{yy}^2
+ 2 \delta \tu_{xy}^2 ) , \nonumber \\
& & + 2 \mu \left([\delta \tu_{xz} -(t/\mu) Q_{xz}]^2 + [\delta
\tu_{yz} - (t/\mu) Q_{yz}]^2 \right) , \nonumber
\end{eqnarray}
where the coefficients ${\overline B}_a$ are evaluated in Appendix\
\ref{app:B}.  This free energy is manifestly invariant under arbitrary
rotations in $S_R$ because it is a function of the strain $\tu_{ij}$
only.  However, its invariance in the $S_T$ is restricted to
infinitesimal rotations in $O_T$ because we only used the harmonic
free energy to integrate over ``massive" modes.  Because underlying
$O_T$ invariance of the nematic state forbids ``massive'' terms in
$Q_{xz}$ and $Q_{yz}$, integration over them also eliminates strains
$v_{xz}$ and $v_{yz}$ from the resulting elastic free energy, which,
as anticipated takes the form identical to that in Eq.\ (\ref{funi2}).
Such symmetry-enforced vanishing of an elastic constant (here $C_5$)
is mathematically closely related to the well-known Anderson-Higgs
mechanism in gauge theories.\cite{ChaLub95}

The terms involving $Q_{xz}$ and $Q_{yz}$ are interesting because
they determine the energy cost of rotating the director away from
the direction of uniaxial stretch.  When we convert to the strain
variables of the stretched state using Eq.\ (\ref{eq:dv}) and the
expressions, Eq.\ (\ref{Lambda2}), for $\Lambda_{0\perp}$ and
$\Lambda_{0||}$ in terms of $S$, we obtain to lowest order in
$\delta \nv = \nv - \nv_0 $
\begin{eqnarray}
f_{\rm el}^{v,n}& = & \case{1}{2} C_1 \eta_{Szz}^{\prime 2} + C_2
\eta_{Szz}(\eta'_{Sxx} + \eta'_{Syy}) \nonumber\\
& & + \case{1}{2} C_3 (\eta'_{Sxx} + \eta'_{Syy})^2 + C_4
(\eta_{Sxx}^{\prime 2} +
\eta_{Syy}^{\prime 2} + 2 \eta_{Sxy}^{\prime 2} )\nonumber\\
& &+\case{1}{2}\mu'\sum_{a=x,y}\left[\eta'_{Saz} - \beta(\delta
n_a-\eta_{Aza}') \right]^2 ,
\label{eq:felv-n}
\end{eqnarray}
where
\begin{equation}
\beta = {(r-1)\over(r+1)}
\label{beta}
\end{equation}
and elastic constants $C_a$ are related to the constants
${\overline B}_a$ via $C_1 = \Lambda_{0||}^4 {\overline B}_1$, $C_2
= \Lambda_{0||}^2\Lambda_{0\perp}^2 {\overline B_2}$, $C_3 =
\Lambda_{0\perp}^4 {\overline B}_3$, $C_4 = \Lambda_{0\perp}^4
{\overline B}_4$, and $\mu' = (r+1)^2 \Lambda_{0\perp}^4\mu$. The
form of this energy is in fact the most general one, and we will
derive it again in the next subsection after we have derived its
nonlinear generalization. It is exactly the form obtained by
Olmsted\cite{Olmsted94} following de Gennes\cite{deGennes} and
Bladon, Terentjev, and Warner\cite{WarBla94}. It shows clearly
how the director can relax locally to $\delta n_a = \eta'_{Aza} +
\beta^{-1} \eta_{Saz}$ to eliminate any dependence of the free
energy on $\eta'_{Saz}$, i.e., to make $C_5=0$.

\subsection{Theory with strain and director}

We have just seen how the development of nematic order
characterized by $Q_{ij}$ leads to a stretched nematic elastomer
with a soft elasticity.  The formulation in terms of $Q_{ij}$ is
well suited to a description of the transition from the isotropic
to the nematic state.  Deep in the nematic phase, the theory that
best captures the effects of long-wavelength strains and
variations in the direction of nematic order is one expressed in
terms of strain and the nematic director $\nv$ only, i.e., one in
which fluctuations in $S$ and in the biaxial part of $Q_{ij}$ are
integrated out.  This theory, like others we have discussed must
be invariant under both rotations in $S_R$ and under simultaneous
rotations of $v_{ij}$ and $\nv$ in $S_T$.

To construct a fully rotationally invariant theory deep in the nematic
phase, it is convenient to introduce a local coordinate system defined
by the orthonormal triad $\{\ev^1 , \ev^2 , \ev^3\equiv\nv\}$
consisting of the local director $\nv$ and two vectors $\ev^1$ and
$\ev^2$ perpendicular to $\nv$.  These vectors satisfy
\begin{mathletters}
\begin{eqnarray}
\ev^\mu \cdot \ev^\nu& = &\delta^{\mu\nu},\\
\sum_{a=1,2}e^a_i e^a_j& = &\delta^\perp_{ij} \equiv \delta_{ij}
- n_i n_j .
\label{eq:basis}
\end{eqnarray}
\end{mathletters}
In what follows, we will adopt a notation in which Greek indices $\mu$
and $\nu$ will run over $1$ to $3$, and Roman indices $a$ and $b$ will
run from $1$ to $2$, i.e., over the subspace transverse to $\nv$. The
left strain tensor $\vm$ can always be expressed in terms of its
components in this basis:
\begin{mathletters}
\begin{eqnarray}
v_{ij} &= &v^{\mu\nu} e_i^\mu e_j^\nu,\\
       & = & v_{||} n_i n_j + v_{\perp}^{ab} e_i^a e_j^b +
v_{||\perp}^a(n_i e_j^a + e_i^a n_j ),
\label{eq:v-ne}
\end{eqnarray}
\end{mathletters}
where
\begin{equation}
v^{\mu \nu} = e_i^\mu v_{ij} e_j^\nu
\label{eq:vmunu}
\end{equation}
and
\begin{mathletters}
\begin{eqnarray}
v_{||}& = &n_i v_{ij} n_j ,\\
v_{||\perp}^a &=& n_i v_{ij} e_j^a \\
v_{\perp}^{ab} & = & e^a_i v_{ij}e^b_j .
\label{eq:com-v}
\end{eqnarray}
\end{mathletters}
The components $v^{\mu\nu}$ are invariant under rotations in $S_R$
because $v_{ij}$ is invariant under $O_R$ by construction.  They are
also invariant under simultaneous rotations of both ${\bf R}$ and the
triad $\{\ev^\mu \}$ in $S_T$, i.e., they maintain their same
numerical value, under simultaneous rotations of $v_{ij}$ and the
basis $\{\ev^1,\ev^2,\nv\}$.

A gel whose anisotropic state forms via spontaneous symmetry
breaking from the isotropic phase has no preferred or imposed
directions, and the elastic free energy will depend only on
$v_{||}$, $v_{\perp}^{ab}$, and $v_{||\perp}^a$.  Furthermore,
this free energy cannot depend on the arbitrary choice of the
vectors $\ev^1$ and $\ev^2$ in the plane perpendicular to $\nv$,
and it will be a function only of $v^{\mu\nu}$ in the
combinations $v_{||}$, $v_{\perp}^{aa}$, $v_{||\perp}^a
v_{||\perp}^a$, and $v_{\perp}^{ab}v_{\perp}^{ab}$. Since linear
terms proportional to $v_{||}$ and $v_{\perp}^{aa}$ are present
in the anisotropic phase, it will be characterized by a
nonvanishing equilibrium strain $v_0^{\mu\nu}$ with components
$v_{0||}$ and $v_{0\perp}^{ab}$.  If the equilibrium director is
$\nv_0$, then such a uniaxially distorted state is characterized
by the equilibrium strain
\begin{eqnarray}
v_{0ij}& = & v_{0||}n_{0i}n_{0j} + v_{0\perp}(\delta_{ij} - n_{0i}
n_{0j}) , \nonumber \\
& \equiv & \case{1}{2}[G_{ij} ( \nv_0 ) - \delta_{ij}] ,
\label{eq:v0}
\end{eqnarray}
where $v_{0||} = (\Lambda_{0||}^2 -1)/2$, $v_{0\perp} =
(\Lambda_{0\perp}^2-1)/2$ and
\begin{equation}
G_{ij} ( \nv ) = \Lambda_{0||}^2 n_i n_j + \Lambda_{0\perp}^2
(\delta_{ij} - n_i n_j) .
\label{eq:Gij}
\end{equation}

Away from equilibrium, the free energy can be expanded in the
deviations
\begin{equation}
\delta v^{\mu\nu} = v^{\mu\nu} - v_0^{\mu\nu}
\end{equation}
of the strain from its equilibrium value. To harmonic order in
these deviations, we have
\begin{eqnarray}
\delta f_w & = & \case{1}{2} C_1 w_{||}^2 +C_2
w_{||}w_{\perp}^{aa} +\case{1}{2} C_3 (w_{\perp}^{aa})^2
\nonumber \\
& & + C_4 w_{\perp}^{ab}w_{\perp}^{ab} + C_5 w_{||\perp}^a
w_{||\perp}^a ,
\label{eq:fw}
\end{eqnarray}
where the rescaled invariant strains are
\begin{eqnarray}
w_{||} & = & \Lambda_{0||}^{-2} (v_{||} - v_{0||}) \nonumber
\\
w_{\perp}^{ab}& = &\Lambda_{0\perp}^{-2}(v_{\perp}^{ab} -
v_{0\perp}^{ab}) \nonumber\\
w_{||\perp}^a & = & {\overline\Lambda}_{0}^{-2}(v_{||\perp}^a -
v_{0||\perp}^a) ,
\label{eq:wstrain}
\end{eqnarray}
with ${\overline \Lambda}_0^2 = (\Lambda_{0||}^2 +
\Lambda_{0\perp}^2)/2$.  The prefactors in these definitions have
been chosen to produce, as we shall see, the simplest linearized
version of the harmonic free energy. Even though this free energy
is only an expansion about a uniaxially-distorted equilibrium
state, unlike the free energies [Eq.(\ref{funi2})] expressed in
terms of strain alone that we considered in Sec.\
\ref{strainonly}, it is completely invariant with respect to
arbitrary rotations in {\em both} $S_R$ {\em and} $S_T$.

The equilibrium values $v_{0||}$ and $v_{0\perp}^{ab}$ do not
depend on the particular direction of $\nv$.  If $\{\ev^\mu \} =
\{\ev_0^\mu \}$, then $v_0^{\mu\nu} =e_{0i}^\mu v_{0ij}e_{0j}^\nu$.
Away from equilibrium defined by $\nv_0$, $v_{ij} = v_{0ij} +
\delta v_{ij}$ where $v_{0ij}$ is given by Eq.\ (\ref{eq:v0}) and
$\delta v_{ij}$ by Eq.\ (\ref{eq:dveta}).   Thus
\begin{equation}
\delta v^{\mu \nu} = (e_i^\mu e_j^\nu - e_{0i}^\mu e_{0j}^\nu )
v_{0ij} + e_i^\mu \delta v_{ij} e_j^\nu ,
\end{equation}
and we have
\begin{mathletters}
\begin{eqnarray}
w_{||} & = &  \Lambda_{0||}^{-2} n_i \delta v_{ij} n_j -
\case{1}{2}(1-(1/r))[1 - (\nv \cdot \nv_0)^2] \nonumber\\
& \approx & n_{0i}\eta'_{ij}n_{0j} = \eta'_{zz} \\
w_{\perp}^{aa} & = & \Lambda_{0\perp}^{-2}
\delta^\perp_{ij}\delta v_{ij}
+ \case{1}{2}(r-1) [1-(\nv\cdot \nv_0)^2] \nonumber\\
& \approx & \delta^{0\perp}_{ij} \eta'_{ij} = \eta'_{xx} +
\eta'_{yy} \\
w_{||\perp}^a w_ {||\perp}^a & =& {\overline\Lambda}_{0}^{-4}
n_{i}n_{k}\delta^{0\perp}_{jl}\delta
v_{ij}\delta v_{kl} \nonumber\\
& & + 2 \beta{\overline\Lambda}_{0}^{-2} (\nv\cdot\nv_0)n_i\delta
v_{ij}\delta^\perp_{jk}n_{0k} \nonumber \\
&  & + \beta^2(\nv\cdot\nv_0)^2[1-(\nv\cdot\nv_0)^2] \nonumber \\
&\approx& \sum_l\left[\eta'_{Szl}-\beta
(\delta n_l - \eta'_{Alz})\right]^2 \\
w_{\perp}^{ab}w_{\perp}^{ab} & =& \Lambda_{0\perp}^{-4}
\delta^\perp_{ik}\delta^\perp_{jl}\delta v'_{ij}
\delta v'_{kl} \nonumber\\
& & + (r-1) n_{0i}n_{0j}\delta^\perp_{ik} \delta^\perp_{lj} \delta
v_{kl} \nonumber\\
& & + \case{1}{4} (r -1)^2 [1 -
(\nv\cdot\nv_0)^2] \nonumber\\
& \approx & \delta_{ik}^{0\perp} \delta_{jl}^{0\perp}
\eta'_{Sij}\eta'_{Skl} = \eta_{xx}^{\prime 2} + \eta_{yy}^{\prime
2} + 2 \eta_{Sxy}^{\prime 2} ,
\end{eqnarray}
\end{mathletters}
where $\beta$ is defined in Eq.\ (\ref{beta}). The energy $\delta
f_w$ is characterized by the five elastic constants $C_a$ and the
stretching ratio $r$, which has the same value in every one of
the nonlinear strains.

Alternative but equivalent expressions for the strains in Eq.\
(\ref{eq:wstrain}) are useful and elegant.  The components of the
equilibrium strains $v_0^{\mu\nu}$ have the same value if the basis
$\{\ev_0^{\mu}\}$ is transformed to the basis $\{\ev^{\mu}\}$ provided
the director $\nv_0$ in $G_{ij}(\nv_0)$ [Eq.\ (\ref{eq:Gij})] is
transformed to $\nv$. Thus we have
\begin{eqnarray}
v_0^{\mu\nu} & = & \case{1}{2}(e_{0i}^\mu G_{ij}( \nv_0 )
e_{0j}^\nu - \delta^{\mu\nu} ) \nonumber \\
& = &\case{1}{2}(e_{i}^\mu G_{ij}( \nv ) e_{j}^\nu -
\delta^{\mu\nu} ) ,
\end{eqnarray}
and from Eqs.\ (\ref{eq:strain2}), (\ref{eq:lambda'}), and
(\ref{eq:vmunu})
\begin{equation}
v^{\mu\nu} = \case{1}{2} [e_i^\mu \Lambda_{ij}' G_{jk} ( \nv_0 )
\Lambda_{kl}^{\prime T} e_l^\nu - \delta^{\mu\nu} ] .
\end{equation}
>From this we obtain
\begin{equation}
\delta v^{\mu\nu} = e_i^\mu V_{ij} e_j^\nu ,
\end{equation}
where
\begin{equation}
\mm{V} = \case{1}{2}\left(\mm{\Lambda}' \mm{G}(\nv_0 )
\mm{\Lambda}^{\prime T} - \mm{G}(\nv)\right),
\end{equation}
and finally
\begin{eqnarray}
v_{||} & = & \Lambda_{0||}^{-2} n_i V_{ij} n_j
\nonumber \\
w_{\perp}^{aa} & = &\Lambda_{0\perp}^{-2} \delta_{ij}^\perp V_{ij}
\nonumber \\
w_{||\perp}^a w_{||\perp}^a & = & \overline{\Lambda}_0^{-4} n_i
n_k \delta_{jl}^\perp V_{ij} V_{kl} \nonumber \\
w_{\perp}^{ab} w_{\perp}^{ab} & = & \Lambda_{0\perp}^{-4}
\delta^\perp_{ik}\delta^\perp_{jl} V_{ij} V_{kl} .
\end{eqnarray}

\subsection{Crosslinking in the Nematic Phase}
\label{sec:nemcross}

If an elastomer is crosslinked in the nematic rather than the
isotropic phase, the memory of the anisotropy of the state, with a
uniaxial direction $\nv_0$, at the time of crosslinking is locked in,
and full $O_R$ invariance of $S_R$ is reduced down to $D_{\infty h}$
symmetry. If coupling to nematic order is turned off, the system will
be characterized by a uniaxial elastic energy of the form of Eq.\
(\ref{funi}) with five elastic constants in general. (Turning off this
coupling is not as unphysical as it may seem. This is precisely what
is done in treatments of plastic crystals consisting of anisotropic
molecules such as $N_2$.) This part of the elastic energy is a
function of $u_{ij}$ and is invariant under rotations in $S_T$.  It is
also invariant under simultaneous rotations of $\nv_0$ and $\xv$ in
$S_R$ and under operations on $\xv$ in $D_{\infty h}$ at fixed
$\nv_0$.  Couplings to the nematic order parameter $\mm{Q}$ must be
invariant under simultaneous rotations of $\Rv$ and $\mm{Q}$ in $S_T$
and under simultaneous rotations of $\xv$ and $\nv_0$ in $S_R$.  The
simplest couplings linear in $\mm{Q}$ are of the form
\begin{equation}
f_C^N = - \Tr \mm{\Lambda}\ \mm{h}\ \mm{\Lambda}^T \mm{Q} -
2\beta\Tr\mm{v} \mm{Q} ,
\label{eq:fCN}
\end{equation}
where $h_{ij} = h n_{0i} n_{0j}$ and, as before,
$\mm{\Lambda}\mm{\Lambda}^T = \mm{\delta} + 2 \mm {v}.$ The first term
in this energy reduces to $-h n_{0i} Q_{ij} n_{0j}$ and favors
alignment of principle axes of $\mm{Q}$ along $\nv_0$ in the absence
of deformation, when the deformation tensor $\mm{\Lambda}$ is the unit
tensor.

The generalization of Eq.\ (\ref{eq:fw}) to systems crosslinked in
the nematic phase is fairly complicated.  It cannot be expressed
in terms of the strain $\delta v_{ij}$ alone;  it can only be
expressed in terms of the more fundamental non-symmetrized strains
$\eta_{ij}$.  However, the major effect of crosslinking in the
nematic phase is to to make $\nv_0$ a preferred direction with an
energy cost to rotate away from that direction, which can be
described by the addition of  a term $ - h (\nv_0\cdot \nv)^2$ to
Eq.\ ({\ref{eq:fw}) to lowest order in $\eta_{ij}$.

\section{Neoclassical Theory of Elastomers}

So far we have described liquid-crystal gels in terms of nonlinear
strains, rotationally invariant in either $S_R$ or in $S_T$, relative
to some equilibrium reference state, and we have focussed on those
properties that result from the spontaneous broken rotational symmetry
of the nematic state. We have treated the elastic constants in our
model free energy as phenomenological parameters to be determined
experimentally.  To date, experimental realization of
liquid-crystalline elastomers are cross-linked liquid-crystalline
polymers.  They are rubbers with orientational degrees of freedom of a
liquid crystal, and their elastic properties over a very wide range of
strains can be described quantitatively by a generalization of the
classic theory of rubber elasticity\cite{BlaTer94}.  This is a
semi-microscopic theory in which the origin of shear moduli is the
reduction of entropy arising from constraining polymers to pass
through cross-linking points. In this section, we will show that this
theory, when expressed in terms of non-linear strains, is equivalent
to those discussed in preceding sections of this article.

In the simplest version of the neoclassical theory, polymer
segments between crosslinks are viewed as independent random-coil
polymers of length $L$.  In the anisotropic environment induced
by the nematic order, the effective step lengths parallel and
perpendicular to the direction of nematic order are different,
and mean-square end-to-end displacement is characterized by an
anisotropic step-length tensor,
\begin{equation}
\mm{l} = l \mm{g} ,
\end{equation}
where $l$ is a length and $\mm{g}$ is a unitless tensor,
reflecting system anisotropy, whose form will be discussed in
different contexts below.  The probability that the two ends of a
single chain are separated by $\Rv$ is
\begin{equation}
P(\Rv) = \left[{\det\mm{l}^{-1}\over (2 \pi L/3)^3}\right]
\exp\left(-{3 \over 2 L} R_i \l^{-1}_{ij} R_j \right) .
\label{prv}
\end{equation}
The free energy per chain is $f_{\rm chain} = - T \ln P(\Rv)$.  Now
assume that the separation $\Rv$ was produced by an affine
transformation from some initial state with separation $\Rv_0$ such
that $R_i = \Lambda_{rij}R_{0j}$, where $\mm{\Lambda}_r$ is the
deformation tensor relative to the initial state. (Later we will
introduce a new reference state and use the symbol $\mm{\Lambda}$ to
denote deformations relative to that state.)  The free energy per
chain of the entire elastomer is then $f_{\rm chain}(\Rv_0)$ averaged
over all separations $\Rv_0$ of the initial state, which we assume
consists of random-walk chain segments characterized by a step-length
tensor $\mm{l}_0=l_0 \mm{g}_0$ and a probability distribution given by
Eq.\ (\ref{prv}) with $\mm{l}$ replaced by $\mm{l}_0$. The initial
state may be viewed as the state at the time of crosslinking.  Thus,
if the system is crosslinked in the isotropic state, $\mm{l}_0$ will
be an isotropic tensor; if it is crosslinked in the nematic state at
some temperature $T$, the degree of anisotropy of $\mm{l}_0$ will
reflect the degree of nematic order at that temperature. The free
energy density relative to the initial state is thus
\begin{equation}
f_{\rm ch} = \case{1}{2} n T (\Tr \mm{\Lambda}_r \mm{l}_0
\mm{\Lambda}_r^T \mm{l}^{-1} - \ln \det \mm{l}_0 \mm{l}^{-1}) ,
\label{eq:fch}
\end{equation}
where $n$ is the volume density of chain segments. This purely
entropic free energy, whose ground state is the collapsed state with
$\mm{\Lambda}_r =0$, cannot alone provide a complete description of
the elastic properties of an elastomer.  It must be supplemented with
some treatment of the short-range enthalpic forces that prevent
collapse to infinite density.  Merely imposing the incompressibility
constraint, $\det \mm{\Lambda}_r=1$, is sufficient to provide a very
good description of dense nearly, incompressible systems.  We will
take a phenomenological approach in which there is an energy cost,
measured by a compression modulus $B_r$, arising from deviations of
$\det\mm{\Lambda}_r$ from $1$:
\begin{equation}
f_{B} = \case{1}{2}B_r (\det \mm{\Lambda}_r -1 )^2 .
\end{equation}
Our complete neoclassical energy density is thus $f= f_{\rm
ch} + f_{B}$.

An important feature of this model is that it depends on
$\mm{\Lambda}_r$ only via the combination $\mm{\Lambda}_r
\mm{g}_0\mm{\Lambda}_r^T$ because the determinant of a product of
tensors is the product of the determinants. Thus, it is convenient
to analyze this model in terms of $\mm{\Lambda} = \mm{\Lambda}_r
\mm{g}_0^{1/2}$, the strain tensor relative to the isotropic state
obtained by rescaling lengths via $\mm{g}_0^{1/2}$.
\cite{MaoWar00} Our model is thus
\begin{eqnarray}
f & = & \case{1}{2} nT (l_0/l) {\rm Tr} \mm{\Lambda}
\mm{\Lambda}^T
\mm{g}^{-1} + \case{1}{2} nT \ln \mm{g}_0 \mm{g}^{-1} \nonumber \\
& & + \case{1}{2}B_r [(\det
\mm{\Lambda}\mm{\Lambda}^T/\det\mm{g}_0)^{1/2} -1]^2 ,
\end{eqnarray}
where $n$ is the volume density of chain segments.

We will now analyze two version of this model: one appropriate to the
description of the $IN$ transition and one appropriate to systems deep in
the nematic phase.  We begin with the $IN$ transition.  In this case, we
take
\begin{eqnarray}
\mm{g}^{-1} & = & \mm{\delta} - \alpha \mm{Q} \nonumber \\
\mm{g}_0 & = & (\mm{\delta}  - \alpha \mm{Q}_0 )^{-1} ,
\end{eqnarray}
where $\mm{Q}_0$ is the value of $\mm{Q}$ at the time of
crosslinking. We could have taken $\mm{g}$ rather than
$\mm{g}^{-1}$ proportional to $\mm{Q}$.  Since we are interested
in small $\mm{Q}$, there is little difference between the two
choices. Our goal is to recast $f$ in terms of the left strain
tensor $\mm{v}$ to obtain a free energy of the form of Eq.\
(\ref{eq:fel-Q}). We begin by finding the equilibrium strain
tensor $\mm{\Lambda}_0$ when $\mm{Q} = 0$.  Since there is no
anisotropy when $\mm{Q}=0$, we have $\mm{\Lambda}_0 =\Lambda_0
\delta_{ij}$.  A straightforward minimization of $f$ with respect
to $\Lambda_0$ when $\mm{Q}=0$ yields the equation of state
\begin{equation}
nT \Lambda_0^2 +B_r(\gamma_0 -1)\gamma_0 =0 ,
\label{eq:eqofstate}
\end{equation}
where $\gamma_0 = (\det\mm{\Lambda}_0 \mm{\Lambda}_0^{T} /\det
\mm{g}_0)^{1/2}$. In the incompressible limit $B_r \rightarrow
\infty$, this yields $\Lambda_0 = (\det \mm{g}_0)^{1/2}$.  Setting
$\mm{\Lambda}\mm{\Lambda}^T = \Lambda_0^2(\mm{\delta} + 2 \mm{v}
)$ and expanding $f$ in powers of $\mm{v}$ using
\begin{eqnarray}
{\rm det}^{1/2}[\mm{\delta} + 2\mm{v}] & = &
\exp[\case{1}{2} \Tr \ln (\mm{1}+ 2 \mm{v})] \nonumber \\
&=&  1 + \Tr \mm{v} - [\Tr \mm{v}^2 - \case{1}{2} (\Tr \mm{v})^2]
+ \cdots ,
\end{eqnarray}
we obtain
\begin{eqnarray}
\delta f & = & \mu {\rm Tr} \mm{v}^2 + \case{1}{2} B ({\rm Tr}
\mm{v})^2
\nonumber \\
& & - \alpha \mu {\rm Tr} \mm{v} \mm{Q} - \case{1}{2} nT {\rm Tr}
\ln(\mm{\delta} - \alpha \mm{Q} )
\end{eqnarray}
to harmonic order in $\mm{v}$, where $\mu=nT(l_0/l) \Lambda_0^2$
and $B = B_r \gamma_0^2 - \mu$. This energy is identical to
$f_{\rm el} + f_C$ of Eq.\ (\ref{eq:fel-Q}) plus a part depending
on $\mm{Q}$ alone, which can be absorbed into $f'_Q$ [Eq.\
(\ref{eq:fQ'})]. The strain can be integrated out to yield
\begin{eqnarray}
\delta f & = &- \case{1}{4} \alpha^2 nT [(l_0/l)\Lambda_0^2 - 1)]
{\rm Tr }\mm{Q}^2 + O(Q^4) \nonumber \\
&\approx&- {1 \over 4} \alpha^2 nT\Bigg[\left({l_0 \over l}
-1\right) -{l_0\over l}{n T \over B} \nonumber\\
& & + {1 \over 6} \alpha^2 {l_0\over l}\left(1-{nT\over B}\right)
{\rm Tr} \mm{Q}_0^2 \Bigg]{\rm Tr} \mm{Q}^2 .
\end{eqnarray}
The final form of this equation was obtained by solving the
equation of state [Eq.\ (\ref{eq:eqofstate})] for $\Lambda_0$ to
lowest order in $nT/B$ and $\alpha^2$. In the incompressible limit
($B=\infty$) when $l_0/l = 1$, there is no shift in the
coefficient of ${\rm Tr} \mm{Q}^2$ and, thus, no shift in the
limit of metastability of the isotropic phase when the system is
crosslinked in the isotropic phase, but there is a small shift
proportional to ${\rm Tr} \mm{Q}_0^2$ when it is crosslinked in
the nematic phase\cite{WarGel88}. If the system is compressible,
$B\neq \infty$, or if the fundamental step lengths $l_0$ and $l$
are different, then there is a shift in the coefficient of ${\rm
Tr}\mm{Q}^2$ even when the system is crosslinked in the isotropic
phase.

Deep in the nematic phase, biaxial fluctuations are suppressed.
If we assume they are completely frozen out, then the step-length
tensor depends only on the director, and we can take
\begin{equation}
{l}_{ij}^{-1} = l_{\perp}^{-1} [\delta_{ij} + (r^{-1} -1 )n_i n_j
]
\label{eq:l-1}
\end{equation}
and $l_{0ij} = l_{0\perp}[\delta_{ij} + (r-1) n_{0i} n_{0j}]$, where
$r = l_{||}/l_{\perp}$.  Since $\mm{l}_0$ has been scaled away by the
transformation from $\mm{\Lambda}_r$ to $\mm{\Lambda}$, the
equilibrium strain $\mm{\Lambda}_0$ for a given $\nv$ will have
components parallel and perpendicular to $\nv$ and will have the form
of Eq.\ (\ref{eq:Lam0}) with $n_{0i}$ replaced by $n_i$.  As $\nv$
rotates so does $\mm{\Lambda}_0$, but the magnitudes $\Lambda_{0||}$
and $\Lambda_{0\perp}$ do not change. Setting $\delta \mm{v} = 0$ and
minimizing over $\mm{\Lambda}_0$, we find the equations of state
\begin{eqnarray}
& & nT {l_{0\perp}\over l_{\perp}}{1 \over r} \Lambda_{0||}^2 + B_r
(\gamma_0 -1 ) \gamma_0 = 0 , \nonumber \\
& & nT {l_{0\perp}\over l_{\perp}}\Lambda_{0\perp}^2 + B_r (\gamma_0 -
1 ) \gamma_0 = 0 .
\end{eqnarray}
These equations imply $\Lambda_{0||}^2 /\Lambda_{0\perp}^2 = r =
l_{||}/l_{\perp}$ for all $B_r$.  Using $\det( \mm{\delta} + 2
\mm{v}_0 + 2 \delta \mm{v}) = \det\mm{\Lambda}_0 \mm{\Lambda}_0^T
\det(\mm{\delta} + \mm{\Lambda}_0^{-1} \delta \mm{v}
\mm{\Lambda}_0^{-1})$ and expanding in $\delta \mm{v}$, we obtain
\begin{eqnarray}
\delta f_w' & = & \mu (w_{||}^2 + w_{\perp}^{ab} w_{\perp}^{ab} )
+ \case{1}{2}
B ( w_{||} + w_{\perp}^{aa} )^2 \nonumber \\
& & + 2 \mu' w_{||\perp}^a w_{||\perp}^a ,
\label{eq:felas1}
\end{eqnarray}
where $\mu = nT (l_{0\perp} /l_{0||}) \Lambda_{0\perp}^2$, $ B =
B_r \gamma_0^2 - \mu$, and $\mu' = \case{1}{4} \mu ( 2 + r +
r^{-1})$.  This is identical to Eq.\ (\ref{funi2}) with $C_1  =
B+ 2 \mu$, $C_2 = B$, $C_3 =B$, $C_4 = \mu$ and $C_5 = 2 \mu'$.

The free energy $\delta f_w'$ of Eq.\ (\ref{eq:felas1}) has a higher
symmetry than the most general free energy $\delta f_w$ of Eq.\
(\ref{eq:fw}): it has only three rather than the five independent
elastic constants. As a result, certain distortions will have the same
energy in the model that do not have the same energy in the most
general model.  For example, purely dilational and compressional
strains with $\Lambda_{xx}$ and $\Lambda_{zz}$ interchanged will have
the same energy in $\delta f_w'$ but not in $\delta f_w$.  The
simplified form of Eq.\ (\ref{eq:felas1}) resulted from our use of
Eq.\ (\ref{eq:l-1}) for $l_{ij}^{-1}$. In general, $l_{ij}^{-1}$
depends on the full tensor order parameter $Q_{ij} = S(n_i n_j -
\case{1}{3} \delta_{ij} ) + B_{ij}$ where $B_{ij}$ is the biaxial part
of $Q_{ij}$ with components in the plane perpendicular to $\nv$.  Deep
in the nematic phase, fluctuations $\delta S$ in the magnitude of $S$
and in $B_{ij}$ are small. The most general form of $l_{ij}^{-1}$ to
lowest order in $\delta S$ and $B_{ij}$ is
\begin{equation}
l_{ij}^{-1} = l_{\perp}^{-1}[ \delta_{ij} + (r^{-1} - 1) n_i n_j +
a\delta S (n_i n_j - \case{1}{3} \delta_{ij}) + b B_{ij} ] ,
\end{equation}
where $a$ and $b$ are numbers.  The nematic energy has
contributions $\case{1}{2} A_S(\delta S)^2 + \case{1}{2} A_B {\rm
Tr} \mm{B}^2$ in addition to the Frank free energy.  Integrating
out $\delta S$ and $B_{ij}$ from the total free energy will yield
an elastic energy in $w_{||}$, $w_{\perp}^{ab}$, and
$w_{||\perp}^a$ with five independent elastic constants, whose
calculation we leave to the reader.

\subsection{Crosslinking in the Nematic Phase}

There is no qualitative distinction in the simple neoclassical theory
between crosslinking in the nematic and isotropic phases.  In both
cases, the equilibrium phase exhibits the soft elasticity
characteristic of spontaneous breaking of the rotational symmetry of
the isotropic state. Thus, additional physics must be added to the
simple neoclassical model to produce the expected memory of the
anisotropy of the nematic state at crosslinking and the concomitant
destruction of soft elasticity.  There are a number of mechanisms that
will produce this memory.  For the purposes of illustration, we will
consider here only a simple model studied by Verwey and
Warner\cite{VerWar95} in which soft elasticity is destroyed via
randomness in the sequence of rigid and flexible units along polymer
chain segments.  The free energy of this model reduces as expected to
the general form discussed in Sec.\ \ref{sec:nemcross}.

The sequence randomness along the chain causes the coupling parameter
$\alpha$ to be a random variable with average $\langle \alpha \rangle$ and
variance $\langle (\delta \alpha)^2 \rangle$.  The chain energy [Eq.\
(\ref{eq:fch})] must be averaged over $\alpha$, which appears in both
$\mm{g}$ and $\mm{g}_0$.  This average (ignoring the $\det \mm{l}_0
\mm{l}^{-1}$ terms) is
\begin{equation}
\langle f_{\rm ch} \rangle = \case{1}{2} nT (l_0/l){\rm Tr}
\mm{\Lambda}_r \langle \mm{g}_0 \rangle \mm{\Lambda}_r^T \langle
\mm{g}^{-1} \rangle + \delta f_{\rm ch} ,
\end{equation}
where
\begin{eqnarray}
\delta f_{\rm ch} & = & \case{1}{2}(l_0/l)nT {\rm Tr}[ \langle
\mm{\Lambda}_r \mm{g}_0 \mm{\Lambda}_r^T \mm{g}^{-1} \rangle -
\mm{\Lambda}_r \langle \mm{g}_0 \rangle \mm{\Lambda}_r^T \langle
\mm{g}^{-1} \rangle ] \nonumber
\\
& \approx & - \case{1}{2} nT (l_0/l)\langle (\delta \alpha)^2
\rangle {\rm Tr} \mm{\Lambda}_r \mm{Q_0} \mm{\Lambda}_r^T \mm{Q}
\end{eqnarray}
We can now proceed as before.  Let $\mm{\Lambda}_r =\mm{\Lambda}\langle
\mm{g}_0\rangle^{-1/2}$, express $\mm{\Lambda}$ in terms of $\mm{v}$, and
expand in powers in $\mm{Q}_0$.  The result is
\begin{eqnarray}
&&\langle f_{\rm ch} \rangle = - (l_0/l)nT \langle \alpha \rangle
{\rm Tr} \mm{v} \mm{Q} \nonumber \\
&& - \case{1}{2} n T \langle (\delta \alpha)^2 \rangle {\rm Tr}
\mm{\Lambda}\langle \mm{g}_0\rangle^{-1/2}\mm{Q_0} \mm{\Lambda}^T
\langle \mm{g}^T_0\rangle \mm{Q} + ...
\end{eqnarray}
This energy is identical to Eq.\ (\ref{eq:fCN}) which we expected
on general grounds.  In the absence of strain, the second term of
this equation tends to align the principal axis of $\mm{Q}$ along
$\mm{Q}_0$.

\section{Conclusions and Future Directions}

In this mostly pedagogical paper we have formulated a classical
elasticity theory of nematic liquid-crystal gels, carefully
incorporating all underlying symmetries and emphasizing the
distinction between independent target and reference space rotational
symmetries. Our formulation leads to a straightforward demonstration
of the soft elasticity of nematic-gel phases that form via spontaneous
symmetry breaking from an isotropic gel. This soft elasticity is
characterized by the symmetry-enforced vanishing of a shear modulus
and vanishing stress up to critical values of the appropriately
applied strain. These and other predictions that emerge from our
formulation are consistent with earlier predictions of the
neo-classical liquid-crystal rubber
theory\cite{FinKoc81}-\cite{Terentjev99}, which had been very
successful in explaining many beautiful experiments on liquid-crystal
elastomers.

The advantage of our formalism is that it elucidates the origin of the
novel soft elasticity of nematic gels, showing that it is dictated by
general symmetry principles common to {\em any spontaneously}
uniaxially ordered elastic medium and is {\em not} limited to any
specific model of such materials. Thereby, our analysis also
demonstrates a close connection between nematically ordered elastomers
to other well-studied ``soft'' lattices, such as smectics (which by
symmetry include cholesterics), columnar phases of fluid liquid
crystals, and tensionless membranes, where rotational symmetry
(corresponding to an arbitrary choice of smectic layers, columns, and
membrane normal orientations) similarly enforces the vanishing of
specific elastic moduli.  This connection allows us to carry over much
of the insight from those systems to gels. For example it seems likely
that the buckling instability\cite{ClarkMeyer73} in smectic liquid
crystals under extensional strain parallel to layer normals will
provide insight into the stripe instability\cite{VerWar96} of a
nematic elastomer subjected to extensional strain perpendicular to its
anisotropy axis or to the as yet unstudied generalization of this
phenomena to compressional strain parallel to the anisotropy axis.

Our formulation also permits a straightforward incorporation of a
variety of important effects such as spatial variations due, e.g.,
to boundary conditions, ever present thermal
fluctuations\cite{GP}, and local gel
heterogeneity\cite{RTaerogel}, thereby allowing a full
statistical-mechanical treatment of nematic elastomers. Again,
experience with smectics\cite{GP,RTaerogel}, columnar
phases\cite{RTaerogel} of conventional liquid crystals, and the
flat phase of tensionless elastic
membranes\cite{NelsonPeliti,AL,LR} strongly suggests that the
latter two effects will qualitatively modify long scales elastic
properties of nematically ordered gels, leading to phenomena such
as, for example, anomalous elasticity, negative Poisson ratio, and
topological glass order. A connection of liquid-crystal gels to a
large body of work on closely-related systems of conventional
liquid crystals confined in {\em rigid} gels, such as the
aerogel\cite{CrawfordZumer96,RTaerogel}, naturally leads to an
important general question: What role does gel elasticity play in
determining the properties and stability of liquid-crystal phases
confined inside {\em flexible} (as opposed to aerogels)
heterogenous gels, such as e.g.,
aerosils\cite{CrawfordZumer96,Leheny}? With the elastic
formulation presented here we plan to address this question in a
future publication.

Finally, the presented description is also natural for treatment of
fluctuating nematic elastomers membranes\cite{membraneXing}, which
constitute a new universality class of membranes, adding to the
well-studied classes of fluid, hexatic and crystalline
membranes\cite{membrane}. In addition to the richness exhibited by
those systems, we expect new physics associated with the interplay of
the in-plane and undulation nonlinear elasticity, both expected to be
important in elastomer membranes\cite{membraneXing}.  Finally, such
in-plane orientationally-ordered elastic membranes are novel
realizations of anisotropic membranes, predicted to exhibit flat,
tubule, and crumpled phases\cite{RTtubules}, subsequently observed in
Monte Carlo simulations\cite{Bowick}. We plan to explore these and
other phenomena and realizations of orientationally ordered elastomers
in future publications.\cite{membraneXing}

\section{Acknowledgments}
We are extremely grateful to Mark Warner for many helpful suggestions
and a critical reading of an early version of this manuscript.  LR and
XX acknowledge the hospitality of Harvard Physics Department, where
part of this work was done. The authors acknowledge generous financial
support for this work from the National Science Foundation under
grants DMR 00-9653 (TCL and RM), MRSEC DMR98-09555 (LR and XX), and
from the A.  P. Sloan and David and Lucile Packard Foundations (LR).

\appendix
\section{Review of Nematic energy}
\label{app:A}
In this appendix, we will review standard treatments of the
isotropic-to-nematic transition, principally to establish notation.
We introduce a complete set of orthonormal symmetric-traceless tensors
$I^{\alpha}_{ij}$ satisfying $\Tr \mm{I}^{\alpha} \mm{I}^{\beta} =
\delta^{\alpha \beta}$:
\begin{eqnarray}
\mm{I}^0 & = &\sqrt{2 \over 3}\left(
\begin{array}{ccc}
    -\case{1}{2} & 0 & 0 \\
    0 & -\case{1}{2} &0 \\
    0 & 0 & 1
\end{array}
\right) ,\qquad
\mm{I}^1 = {1 \over \sqrt{2}} \left(
\begin{array}{ccc}
    1 & 0 & 0 \\
    0 & -1 & 0 \\
    0 & 0 & 0
\end{array}
\right) \nonumber \\
\mm{I}^2 & = &{1 \over \sqrt{2}} \left(
\begin{array}{ccc}
    0 & 1 & 0 \\
    1 & 0 & 0 \\
    0 & 0 & 0
\end{array}
\right) ,\qquad
\mm{I}^3  = {1 \over \sqrt{2}} \left(
\begin{array}{ccc}
    0 & 0 & 1 \\
    0 & 0 & 0 \\
    1 & 0 & 0
\end{array}
\right) , \nonumber\\
\mm{I}^4 & = &{1 \over \sqrt{2}} \left(
\begin{array}{ccc}
    0 & 0 & 0 \\
    0 & 0 & 1 \\
    0 & 1 & 0
\end{array}
\right) .
\end{eqnarray}
Any symmetric-traceless tensor can be expressed as a linear
combination of these matrices: $Q_{ij} = \sum_{\alpha=0}^4
Q_{\alpha} I^{\alpha}_{ij}$, where $Q_{\alpha} = \Tr \mm{Q}
\mm{I}^{\alpha}$. Thus,
\begin{eqnarray}
Q_0 & = & \sqrt{\case{2}{3}} [Q_{zz} - \case{1}{2}(Q_{xx} +
Q_{yy})], \qquad Q_1 = \case{1}{\sqrt{2}}(Q_{xx} - Q_{yy}),
\nonumber \\
Q_2 & = & \sqrt{2} Q_{xy}, \qquad Q_3 = \sqrt{2}Q_{xz}, \qquad
Q_4 = \sqrt{2} Q_{yz}.
\label{Qalpha}
\end{eqnarray}
With $\nv^0$ along the $z$ axis, $I^0_{ij} = \sqrt{3/2}(n^0_i
n^0_j - (1/3) \delta_{ij})$.  In uniaxial nematic phase,
$Q^0_{ij} = S (n^0_i n^0_j - (1/3) \delta_{ij}) = Q_0
I^0_{ij}$ and $Q_0 = \sqrt{2/3} S$.  The Landau-de Gennes
free energy for a nematic is
\begin{eqnarray}
& &f_Q =\case{1}{2} r_Q {\rm Tr}\mm{Q}^2 -
w_3 {\rm Tr} \mm{Q}^3 + w_4 ({\rm Tr}\mm{Q}^2)^2, \nonumber \\
&&= \case{1}{2} r_Q \sum_{\alpha} Q_{\alpha}^2 + w_4
\left(\sum_{\alpha} Q_{\alpha}^2 \right)^2 \nonumber \\
& & - w_3 \left[\case{1}{2}\sqrt{\case{2}{3}} Q_0^3
-\sqrt{\case{3}{2}} Q_0 (Q_1^2 + Q_2^2) +
\case{1}{2}\sqrt{\case{3}{2}}Q_0(Q_3^2 + Q_4^2)
\right]\nonumber \\
& & - \case{1}{2\sqrt{2}}w_3[3 Q_1 (Q_2^2 - Q_3^2) + 6 Q_2 Q_3
Q_4] .
\end{eqnarray}
Minimization with respect to $Q_0$ yields the equation of
state
\begin{equation}
r_Q Q_0 + 4 w_4 Q_0^3- \sqrt{\case{3}{2}} w_3 Q_0^2 = 0 .
\end{equation}
Then expansion to second order in deviations $\delta
Q_{\alpha} = Q_{\alpha} - Q_{\alpha}^0$ yields
\begin{equation}
\delta f = \case{1}{2}A_1 (\delta S)^2 + \case{1}{2} A_2 (Q_1^2 +
Q_2^2) .
\end{equation}
where
\begin{eqnarray}
A_1  & = & \case{2}{3}(r_Q + 8 w_4 S^2 - 2 w_3 S)
\nonumber \\
A_2 & = & 3 w_3 S
\label{AdfQ}
\end{eqnarray}
As anticipated from the underlying rotational invariance, there are no
terms proportional to $\delta Q_3^2$ or $\delta Q_4^2$.

\section{Evaluation of $B_a$}
\label{app:B}
In this appendix, we outline the algebraic steps between Eqs.\
(\ref{dfel-Q}) and (\ref{eq:fv1}).  We need to integrate over $\delta
S$ and $Q_1$ and $Q_2$.  Since these variables appear only to
quadratic order, the integration is trivial and yields
\begin{eqnarray}
\delta f_{u-v} &= &\case{1}{2} B_1 ({\rm Tr} \delta \mm{u} )^2 +
\mu_2 (v_1^2 + v_2^2) \nonumber \\
& & + \mu_0\delta v_0^2 - \gamma\delta v_0 {\rm Tr}\delta \mm{u} ,
\end{eqnarray}
where
\begin{eqnarray}
B_1 & = &  B - {16 s^2 S^2 \over 9 A'}
\nonumber\\
\mu_0 & = & \mu - {4 t^2 \over 3 A'} \nonumber\\
\mu_2 & = & {\mu A_2  \over A_2 + (2 t^2/ \mu)} \nonumber \\
\gamma & = & - 4 \left({2\over 3}\right)^{3/2} {s t S\over A'}
\end{eqnarray}
where
\begin{equation}
A' = A_1 +{9 s^2 \over 4 B} S^2 + {3 t^2 \over \mu} .
\end{equation}
Setting ${\rm Tr}\delta \mm{u} = (\delta v_{xx} + \delta v_{yy} +
\delta v_{zz} )$, replacing $v_0$, $v_1$, and $v_2$ with expressions
obtained from Eqs.\ (\ref{Qalpha}) (with the tensor $\mm{Q}$ replaced
by $\delta \mm{v}$) and including the $\delta v_{\alpha} - (t/\mu)
\delta Q_{\alpha}$ with $\alpha = 3,4$ terms in Eq.\ (\ref{dfel-Q}),
we obtain Eq.\ (\ref{eq:fv1}) with
\begin{eqnarray}
{\overline B}_1 & = &B_1 + {4 \over 3} \mu_0 - 2 \sqrt{2\over 3}
\gamma \nonumber\\
{\overline B}_2 & = & B_2 -{2 \over 3} \mu_0 -{1\over 2} \sqrt{2
\over 3} \gamma \nonumber \\
{\overline B}_3 & = & B_1 - \mu_2 +{1\over 3} \mu_0 + \sqrt{2
\over 3} \gamma \nonumber \\
{\overline B}_4 & = & \mu_2 .
\label{overB}
\end{eqnarray}

\section{Linearized limits of Eulerian and Lagrangian Elasticity}
\label{app:C}

It is often the case that a linearized theory of elasticity, in which
nonlinear strains are replaced by their linearized limits and only
terms to harmonic order in these linearized strains are included in
the free energy, provides an adequate description of elastic
distortions.  It is, therefore, interesting to see how this linearized
limit is reached.  It turns out that this limit can be taken more
cleanly in the Eulerian picture in which the displacement field is a
part of the phase of a mass-density wave rather than the Lagrangian
picture in which $\uv(\xv)$ is a displacement relative to a reference
configuration. For a further discussion of these two pictures of
elasticity, see Ref.\ \onlinecite{ChaLub95}.  Much of our intuition
about how to construct a linearize theory comes from the Eulerian
picture in which the displacement field is a vector field in space
that obeys the usual rules of transformation of vector fields.  In
this Appendix, we will discuss the linearized limits of Eulerian and
Lagrangian elasticities.

\subsection{Eulerian Elasticity}

In Eulerian elasticity, the displacement field $\uv(\xv)$ is a
vector field in three space.  Like all vector fields that
transform under the same group as space itself, $\uv$ transforms
under a rotation of the whole sample as
\begin{equation}
\uv'( \xv) = U\uv ( \xv ) U^{-1} = \mm{O}\uv ( \mm{O}^{-1} \xv )
\end{equation}
where $U$ is a rotation operator (e.g., quantum mechanical operator)
and $\mm{O}$ is its associated 3$d$ rotation matrix.  Here the prime
indicates the value of the field after the rotation operator is
applied. To leave the system unchanged, $U$ must be an operation in
the point group of the crystal. In the Eulerian picture, $\uv(\xv)$ is
a Goldstone field associated with the broken spatial symmetry of a
crystal.  Thus, strictly speaking the highest-symmetry point group in
three dimensions in the cubic group.  To make contact with our
discussion of gels, we can, however imagine a system in which all
rotations are in the point group. Since $\uv$ is a vector field,
$\partial_i u_j$ is a tensor field that satisfies
\begin{equation}
(\partial_i u_j)' ( \xv ) = U\partial_i u_j U^{-1} =
O_{ik}\partial'_k u_l ( \xv')O_{lj}^{-1} ,
\end{equation}
where $\xv' \equiv \mm{O}^{-1} \xv$, $\partial_i' =
\partial/\partial x'_i$, and as before the prime indicates the value of the
operator after rotation. Alternatively, we can introduce
\begin{equation}
\eta_{ij} = \partial_j u_i ,
\end{equation}
which
\begin{equation}
\mm{\eta}' =\mm{O}\ \mm{\eta}\ \mm{O}^{-1}
\end{equation}
Scalars created from $\mm{\eta}$ or from $\uv$ and its derivatives
are invariant under $U$.  For example,
\begin{equation}
\eta_{ii} = \partial_i u_i, \qquad \eta_{ij} \eta_{ji}
\end{equation}
etc. are scalars under $U$.

The above symmetries and considerations apply to {\em any} vector
field.  The displacement field $\uv$, however, has additional
properties arising from the fact that it is a Goldstone field.  In
particular the systems is invariant under rotation of the
mass-density wave crystal (which is not the same thing as rotating
the whole sample).  The transformation
\begin{equation}
\xv - \uv ( \xv ) \rightarrow \mm{O}(\xv -\uv ( \xv ) )
\end{equation}
rotates the crystal.  Thus, the transformation
\begin{equation}
\uv \rightarrow \uv' ( \xv ) = -\mm{O}(\xv - \uv ( \xv ) ) + \xv
\end{equation}
does not change the energy of the system.  This implies that the
elastic energy will depend only on the Eulerian {\em symmetrized}
strain:
\begin{eqnarray}
u_{ij}^E &= &\partial_i u_j + \partial_i u_j - \partial_i u_k
\partial_j u_k \\
& \approx &\partial_i u_j + \partial_i u_j
\end{eqnarray}
where the final form is its harmonic limit.

Thus, we have two symmetries: (1) symmetries associated with
rotation of the whole sample, and (2) rotations of the lattice.
Invariance with respect to the first requires that the energy
depend only on scalars formed by contracting indices of both
gradients and $\uv$'s.  The second invariance requires that the
energy be a function only of the nonlinear strain $u_{ij}^E$.  The
interesting thing is that {\em both} $u_{ij}^E$ {\em and} its
linearized form transform like tensors under (1), i.e., under $O$.
Thus, contracted tensors of either $\mm{u}^E$ or its linearized
form are scalars under $O$.

\subsection{Lagrangian Elasticity}

In Lagrangian Elasticity, there are as we have discussed two
symmetries: (1) rotations $O_T$ in the target space and (2)
rotation $O_R$ in the reference space. Under these operations, the
displacement vector satisfies
\begin{equation}
\Rv' ( \xv ) = \mm{O}_T \Rv( \mm{O}_R^{-1} \xv )
\end{equation}
Under infinitesimal rotations,
\begin{eqnarray}
O_{Tij} & = &\delta_{ij} + \epsilon_{ijk} \theta_{Tk}\\
O_{Rij} & = & \delta_{ij} + \epsilon_{ijk} \theta_{Rk} .
\end{eqnarray}
and
\begin{eqnarray}
\partial_i u'_j& = &\partial_i' u_j +\epsilon_{ijp} (\theta_{Rp} - \theta_{Tp})
\nonumber \\
&& + \epsilon_{ikp} \partial'_k u_j \theta_{Rp} - \epsilon_{kjp}
\partial'_i u_k \theta_{Tp}
\end{eqnarray}
The energy is invariant under independent rotations through
$\theta_R$ and $\theta_T$.  As we have seen, these invariances are
guaranteed by making the free energy a function only of the fully
contracted $u_{ij}$ or $v_{ij}$ tensors.

Now, let us look at the linearized limit.  Under both rotations, we
have,
\begin{eqnarray}
\partial_i u'_j + \partial_j u'_i & = & \partial'_i u_j +
\partial'_j u_i \nonumber \\
&& +(\epsilon_{ikp} \partial_k' u_j + \epsilon_{jkp} \partial_k' u_i
) \theta_{Rp} \nonumber \\
& & - (\epsilon_{kjp} \partial_i' u_k + \epsilon_{kip} \partial_j'
u_k) \theta_{Tp} .
\end{eqnarray}
Note that to leading order in $u$ and $\theta$, this symmetrized
combination is independent of $\theta_R$ and $\theta_T$ as it
should be.  The terms of order $\theta u$ tell us about the
tensorial rotation properties of the system.  {\em If} $\theta_R=
\theta_T = \theta$, then the symmetrized combination $u_{ij}^S
=\partial_i u_j + \partial_j u_i$ transforms like a tensor, i.e.,
\begin{equation}
\mm{u'}^S = \mm{O} \mm{u}^S \mm{O}^{-1} .
\end{equation}
The linearized strain $\mm{u}^S$ does not, however, transform like
a tensor under independent rotations $\theta_R$ and $\theta_T$.
If, for example, $\theta_R$ is zero, the term proportional to the
product $\theta_{Ri} u_k$ in the transformation of $\mm{u}^S$
depends on both the symmetric and anti-symmetric parts of
$\partial_i u_j$. Only the fully nonlinear strains $u_{ij}$ and
$v_{ij}$ transform like tensors even to linear order in $\theta_R$
or $\theta_T$. We leave it as an exercise to verify this
explicitly. {\em Thus, we cannot use the linearized tensors to
discuss the rotation and tensorial properties of strains in the
Lagrangian language.} We can, however, as discussed above use them
fruitfully in the Eulerian language.  However, once we have
constructed a rotationally invariant Lagrangian energy, we can
replace nonlinear strains by linearized ones to discuss harmonic
elastic fluctuations.


\begin{thebibliography}{10}
\bibitem{comment_slow} By slow we mean a distortion whose wavelength is
much longer than the mesh spacing.
\bibitem{LagrangeE} In Lagrangian elasticity, the fundamental variables are
displacements of mass points parametrized as positions in a
reference material.  In Eulerian elasticity, the fundamental
variables are phases of mass-density waves as a function of
position in space.  See Ref.\ \cite{ChaLub95}, Chap. 6. and App.\
\ref{app:C}
\bibitem{deformations}
The combinations $\mm{\Lambda}^T\mm{\Lambda}$ and
$\mm{\Lambda}\ \mm{\Lambda}^T$ are referred to, respectively, as
the right and left Cauchy-Green deformation tensors\cite{strain}.
\bibitem{invar}
In three dimensions, there are of course only three independent
invariants, which can, for example, be ${\rm
Tr}\mm{\Lambda}\ \mm{\Lambda}^T$, ${\rm
Tr}(\mm{\Lambda}\ \mm{\Lambda}^T)^2$, and
$\det\mm{\Lambda}\ \mm{\Lambda}^T$ or ${\rm
Tr}(\mm{\Lambda}\ \mm{\Lambda}^T)^3$.  All other invariants such as
${\rm Tr}(\mm{\Lambda}\ \mm{\Lambda}^T)^4$ can be expressed in
terms of these three invariants.
\bibitem{incompressible_comment} The novel soft-elasticity physics of
  liquid-crystal gels that we are interested in is contained in $f$.
  To make contact with incompressible rubber-elasticity
  models\cite{WarTer96}, $B\rightarrow\infty$ limit must be taken,
  restricting the remaining volume nonpreserving degree of freedom
  $\Tr\um$ to zero.
\bibitem{dirrelax}
In a more microscopic description, the director must also be
allowed to relax - see refs. \onlinecite{WarBla94},
\onlinecite{VerWar95}, and \onlinecite{Mitchell}.  If director and
strains are not allowed to relax, the materials are ``hard" until
an elastic instability is reached\cite{Mitchell}
\bibitem{comment_C4} As is clear from Eq.\ (\ref{Cs}), within a simple
  quartic model of $\mm{\ut}$, Eq.\ (\ref{eq:f_2}), valid only close to the
  isotropic-uniaxial transition (and even then, only if it is weakly
  first order, so that $\psi$ is small), the elastic constant $C_4$ is
  is strictly positive, precluding a uniaxial-to-biaxial deformation
  transition.  However, a more general free energy density of
  $\mm{\ut}$ (e.g., including higher order nonlinearities in
  $\mm{\ut}$) can be shown to lead to $C_4$ that does go negative,
  thereby triggering the transition to a biaxial distortion.
\end{thebibliography}
\end{document}